\documentclass[aip,jap,reprint]{revtex4-1}

\usepackage{graphicx}% Includes figures
\usepackage{scrextend}% Allows to use one footnote multiple times

\begin{document}

%\preprint{}

\title{Optical spectroscopy on the photo-response in multiferroic BiFeO\textsubscript{3} at high pressure}

\author{F. Meggle}
\affiliation{Experimentalphysik 2, Universit\"at Augsburg, 86159 Augsburg, Germany}

\author{J. Ebad-Allah}
\affiliation{Experimentalphysik 2, Universit\"at Augsburg, 86159 Augsburg, Germany}
\affiliation{Department of Physics, Tanta University, 31527 Tanta, Egypt}

\author{J. Kreisel}
\affiliation{Physics and Materials Science Research Unit, University of Luxembourg, 4422 Belvaux, Luxembourg}
\affiliation{Department Materials Research and Technology, Luxembourg Institute of Science and Technology, 41 Rue du Brill, 4422 Belvaux, Luxembourg}

\author{C. A. Kuntscher}\email{christine.kuntscher@physik.uni-augsburg.de}
\affiliation{Experimentalphysik 2, Universit\"at Augsburg, 86159 Augsburg, Germany}


\date{\today}

\begin{abstract}
The pressure dependence of light-induced effects in single-crystalline BiFeO$_3$ is studied by optical spectroscopy. At low pressures, we observe three light-induced absorption features with energies just below the two crystal field excitations and the absorption onset, respectively. These absorption features were previously ascribed to excitons, possibly connected with the ultra-fast photostriction effect in BiFeO$_3$.
The pressure-induced redshift of the absorption features follows the pressure dependence of the corresponding crystal field excitations and absorption onset, suggesting the link between them. Above the structural phase transition at $P_{\mathrm{c1}}\approx{}3.5$~GPa the three absorption features disappear, suggesting their connection to the polar phase in BiFeO$_3$. The pressure-induced disappearance of the photo-induced features is irreversible upon pressure release.
\end{abstract}


%\keywords{BiFeO$_3$, morphotropic phase boundary, infrared spectroscopy} Is this option still allowed?

\maketitle

\section{Introduction}

Materials with the chemical formula $ABX_3$, where $A$ and $B$ denote cations and $X$ denotes an anion, often deviate from the ideal cubic perovskite structure\cite{Mitchell.2002} (space group $Pm\overline{3}m$) and show distortions leading to a plethora of interesting physical properties such as piezoelectricity, pyroelectricity, (anti-) ferroelectricity, or even multiferroicity.\cite{Li.2015(a), Liu.2017, Hill.2000}
An intensively studied perovskite oxide is bismuth ferrite BiFeO$_3$ (BFO). It crystallizes at ambient conditions in a highly distorted perovskite rhombohedral $R3c$ structure with lattice parameter $a_{\mathrm{rh}}=5.6343~\mathring{\mathrm{A}}$ and $\alpha{}_{\mathrm{rh}}=59.348^\circ$.\cite{Kubel.1990} BFO presents oxygen octahedra $a^-a^-a^-$ anti-phase tilts in Glazer's notation\cite{Glazer.1972}, together with an important displacement of the Bi and Fe cations along the [111]$_{\mathrm{pc}}$ pseudocubic direction.\cite{Kubel.1990} The large cation displacement results from the stereochemically active Bi(6$s^{2}p^{0}$) lone pair\cite{Iniguez.2003, Seshadri.2001}, and leads consequently to an important net ferroelectric polarization of BFO. The high theoretical polarization value of around 90~$\mu{}$m/cm$^2$ (see Ref.\ \citenum{Neaton.2005}) was experimentally confirmed.\cite{Shvartsman.2007, Lebeugle.2007(b)}

Ferroelectricity in perovskite oxides can be explained by an imbalance between Coulomb interactions favoring ferroelectric distortions and short-range repulsion which prefer the undistorted high-symmetry structure.\cite{Cohen.1990, Cohen.1992} % Hybridization of the oxygen 2$p$ and the $B$-site transition metal $d$ orbitals decreases the short-range repulsion.\cite{Posternak.1994}
By applying hydrostatic pressure on a ferroelectric crystal, the short-range repulsions increase faster than the Coulomb interactions, leading to a reduction and even to the disappearance of ferroelectricity in perovskite crystals.\cite{Samara.1975} Interestingly, a report of Kornev et al.\cite{Kornev.2005} predicted the reappearance of ferroelectricity at even higher pressures, which was verified experimentally on the model ferroelectric perovskite PbTiO$_3$.\cite{Janolin.2008}

Accordingly, the transition from the ferroelectric to the paraelectric state is related to a structural phase transition. BFO undergoes multiple structural phase transitions under pressure, where the first phase transition occurs at $P_{\mathrm{c1}}\approx{}3.5$~GPa.\cite{Guennou.2011, GomezSalces.2012, Wu.2018} There are inconsistent reports regarding the crystal structure of BFO above $P_{\mathrm{c1}}$ (see Fig.\ 2 in Ref.\ \citenum{Guennou.2011}), including orthorhombic\cite{Guennou.2011, Mishra.2013, Belik.2009, Kozlenko.2011, Wu.2018, Buhot.2015}, monoclinic\cite{Haumont.2009, Chen.2012(b), Guennou.2011(b)} or a mixture\cite{Zhang.2013, Wu.2018, Guo.2019} of various phases. The corresponding space group for the possible monoclinic symmetry was suggested to be $C2/m$ and the orthorhombic phases were reported to exhibit $Ima2$, $I2cm$, $I2cb$, $Pbam$, $Ibam$, $Cmmm$, $Pna2_1$ or even $P222_1$ symmetry. The space groups $Ima2$, $I2cm$, $I2cb$ and $Pna2_1$ proposed by Guennou et al.\cite{Guennou.2011} and Buhot et al. \cite{Buhot.2015} possess ferroelectric ordering\cite{Shi.2016} and the $Pbam$ structure is by symmetry anti-polar\cite{Kozlenko.2011}. The other reported orthorhombic phases exhibit a non-polar symmetry. In the pressure range 10 -- 12~GPa a structural phase transition to the macroscopically non-polar $Pnma$ phase occurs.\cite{Wu.2018, Guennou.2011, Haumont.2009, Guo.2019}

A very interesting subarea of ferroelectric compounds is their interaction with light, for example, above-band gap voltages, optical control of polarization, photoelectricity, or an enhancement of ferroelectric polarization under light illumination.\cite{Yang.2010, Li.2018, Moubah.2012, Kreisel.2012, Borkar.2018} A still not completely understood mechanism is the so-called photostriction effect, i.e., incident light changes the lateral dimensions of a crystal.\cite{Figielski.1961, Wei.2017(a), Kundys.2010, Wei.2017(b)} Early reports\cite{Uchino.1985, Dingquan.1991} explained photostriction as a superposition of the bulk-photovoltaic and the inverse piezoelectric effect. However, ultra-fast time-resolved x-ray diffraction (XRD) studies on BFO disagree with this classical explanation and claimed in the case of BFO the creation of excitons during light illumination.\cite{Schick.2014} Two recent optical spectroscopy studies\cite{Burkert.2016, Meggle.2019} observed three absorption features on BFO single crystals during laser illumination which are energetically close to the crystal field excitations and the absorption onset. These features were interpreted in terms of excitons. Temperature-dependent measurements\cite{Meggle.2019} suggested a coupling of the light-induced excitons to phonons and potentially also to magnons.

Here, we study the effect of hydrostatic pressure on the photo-induced absorption features in BFO, in order to investigate a potential link between the absorption features and the ferroelectric $R3c$ phase. The pressure-induced phase transition from the polar to a non-polar structure in BFO opens the possibility to gain further information on the mechanism underlying the photostriction effect in BFO.

\section{Experimental Method}
The transmission measurements in the frequency range 8500 -- 18000~cm$^{-1}$ (1.05 to 2.23~eV) were carried out with a Bruker IR-scope coupled to a Bruker Vertex 80v FTIR spectrometer.
A clamp diamond anvil cell (Diacell cryoDAC-Mega) with a culet diameter of 500~$\mu{}$m generated pressures up to 5.3~GPa.
The investigated BFO single crystal was grown by the flux method as described in Refs.\ \citenum{Lebeugle.2007(b)} and \citenum{Haumont.2008}.
We cut a small piece with lateral dimensions of approx.\ 150~$\mu{}$m $\times{}$ 75~$\mu{}$m from the very same BFO single crystal which was used for optical measurements at ambient conditions in Ref.\ \citenum{Burkert.2016}. The single crystal was polished to a thickness of approx.\ 35~$\mu{}$m. The sample is in a multidomain state [see polarized light microscopy image in Fig.\ 1(d) of Ref.\ \citenum{Burkert.2016}]. Nevertheless, the photo-induced changes are representative, since the probing spot was kept constant during the whole pressure cycle. We placed the sample in the hole of a CuBe gasket and used an alcohol mixture (methanol:ethanol=4:1) as pressure transmitting medium since it provides hydrostatic conditions up to 10.5~GPa.\cite{Klotz.2009} For the pressure determination inside the diamond anvil cell (DAC) we used the ruby R-line luminescence shift.\cite{Syassen.2008} In our pressure cycle up to 5.3~GPa the ruby luminescence spectra show symmetric R1 and R2 peaks underpinning the hydrostatic pressure conditions.

We measured the intensities $I_{\mathrm{BFO}}(\nu{})$ and $I_{\mathrm{ref}}(\nu{})$ of the radiation transmitted through the BFO crystal and the pressure transmitting medium in the DAC, respectively. The transmission and absorbance spectra were calculated according to $T(\nu{})=I_{\mathrm{BFO}}(\nu{})/I_{\mathrm{ref}}(\nu{})$ and $A(\nu{})=-\log_{10}{T(\nu{})}$, respectively.

The setup for measuring the photo-induced optical response is similar to the one described in Ref.\ \citenum{Meggle.2019}: We used a blue laser for excitation ($\lambda{}$=473~nm, $E$=2.6~eV, $P$=23.5~mW, polarization ratio larger than 100:1, beam diameter: approx.\ 1.2~mm, cw). A $45^{\circ}$ mirror was fixed below the upper Cassegrain objective of the IR-scope, in order to deflect the laser beam onto the sample, and a longpass filter with cut-off wavelength of $\lambda{}_{\mathrm{cut-off}}$=495~nm was mounted in front of the detector. In addition, we placed a converging lens between the laser and the $45^{\circ}$ mirror, in order to increase the energy density of the laser on the sample. The focused laser spot had a diameter of $\sim$200~$\mu$m leading to an energy density of around 750~mW/mm$^2$. This is orders of magnitudes smaller than the energy densities which were used in former Raman-measurements on BFO.\cite{Himcinschi.2019, Himcinschi.2015, Liang.2019, Kreisel.2011} Accordingly, we can exclude that the laser leads to a temperature increase of the sample during our measurements.

\begin{figure}[t]
\includegraphics[width=0.9\columnwidth]{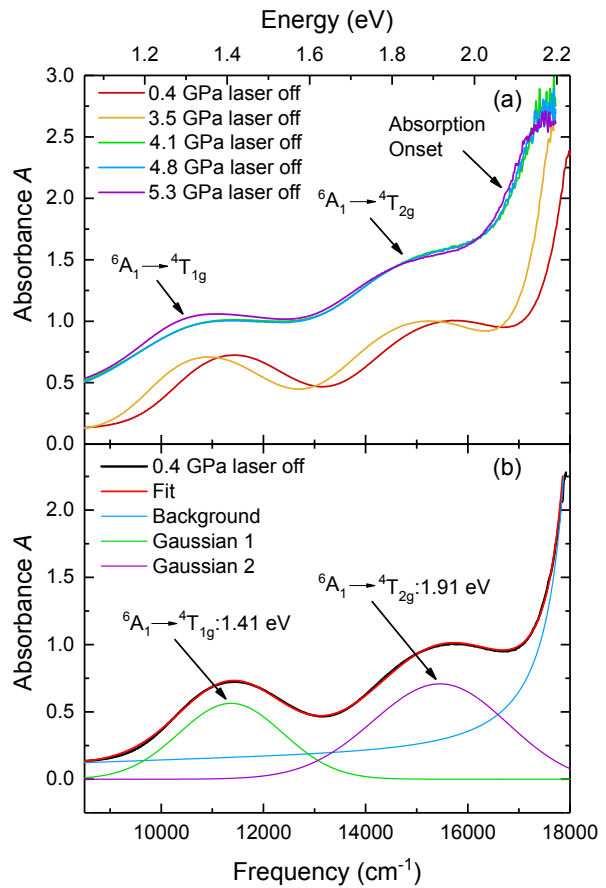}
\caption{(a) Absorbance spectra of the BFO single crystal for selected pressures between 0.4 and 5.3~GPa. (b) Fit of the absorbance spectrum of BFO at 0.4~GPa. The fit contains two Gaussian functions for the crystal field excitations and one Lorentzian term describing the absorption onset.}
\label{fig:absorbance}
\end{figure}

\begin{figure*}
\includegraphics[width=1\textwidth]{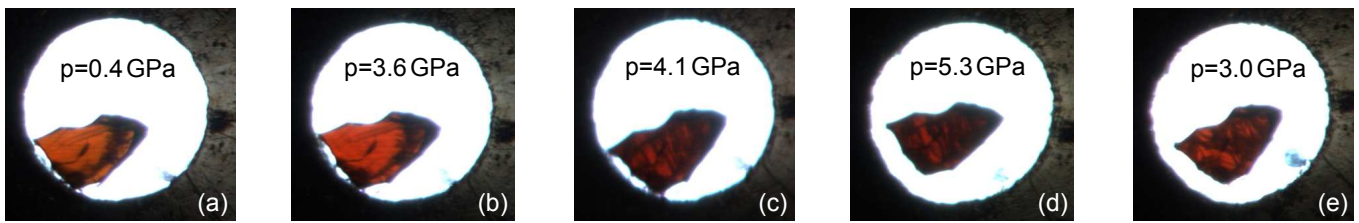}
\caption{Images of the BiFeO$_3$ single crystal at selected pressures between 0.4 and 5.3~GPa. The pictures (a)-(d) were taken during pressure increase, while the photo in (e) was recorded during pressure release.}
\label{fig:pictures_sample_under_pressure}
\end{figure*}

\section{Results and Discussion}

Absorbance spectra of the BFO single crystal for selected pressures between 0.4 and 5.3~GPa without laser illumination are depicted in Fig.\ \ref{fig:absorbance}(a). All spectra show similar characteristics, namely two absorption bands due to $d$-$d$ crystal field excitations
($^{6}\mathrm{A}_{1\mathrm{g}}\rightarrow{}^{4}\mathrm{T}_{1\mathrm{g}}$,
$^{6}\mathrm{A}_{1\mathrm{g}}\rightarrow{}^{4}\mathrm{T}_{2\mathrm{g}}$) and a steep absorption onset at higher energies, consistent with the literature.\cite{Ramachandran.2010,Neaton.2005,GomezSalces.2012,Xu.2009,Kumar.2008,Burkert.2016} At 0.4~GPa the crystal field transitions are located at $1.41$~eV and $1.91$~eV, respectively, which is in fair agreement with previous optical measurements on BFO under pressure.\cite{GomezSalces.2012} With increasing pressure the crystal field excitations and the absorption onset shift to lower energies consistent with previous reports.\cite{GomezSalces.2012}

\begin{figure}[b]
\includegraphics[width=0.9\columnwidth]{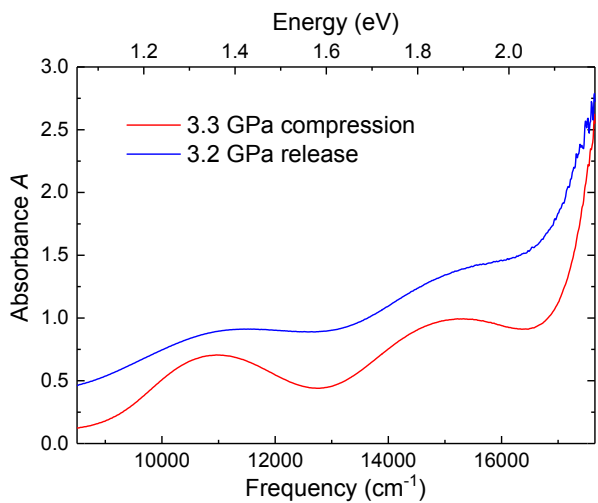}
\caption{Comparison of the absorbance spectra of the BFO single crystal during pressure increase (red spectrum) and release (blue spectrum) without laser illumination.}
\label{fig:decompression}
\end{figure}

Between 3.5 and 4.1~GPa the overall absorbance increases abruptly [see Fig.\ \ref{fig:absorbance}(a)]. Hereby, the sample changes its color from mainly reddish at $P$=3.5~GPa to a predominant black color at 4.1~GPa [see Figs. \ref{fig:pictures_sample_under_pressure}(b) and (c)]. With further pressure increase up to the highest measured pressure (5.3~GPa) only marginal changes occur in the absorbance spectra, where mostly the $^{6}\mathrm{A}_{1\mathrm{g}}\rightarrow{}^{4}\mathrm{T}_{1\mathrm{g}}$ crystal field transition is affected [see Fig.\ \ref{fig:absorbance}(a)]. The color of the sample stays also rather constant between 4.1 and 5.3~GPa [see Figs. \ref{fig:pictures_sample_under_pressure}(c) and (d)]. During pressure release we observe a remarkable non-reversibility of the pressure-induced changes in the absorbance spectrum
(see Fig.\ \ref{fig:decompression}): the overall absorption remains at a higher level which is comparable to the spectra observed above the critical pressure $P_{\mathrm{c1}}$ [see Fig.\ \ref{fig:absorbance}(a)]. By comparing the images shown in Figs.\ \ref{fig:pictures_sample_under_pressure}(b) and \ref{fig:pictures_sample_under_pressure}(e) (pressure increase at 3.6~GPa vs. pressure release at 3.0~GPa) one notices that the sample colors differ from each other. The color of the sample at 3.0~GPa during pressure release is comparable to the color of the sample at 4.1~GPa.

\begin{figure}[b]
\includegraphics[width=1\columnwidth]{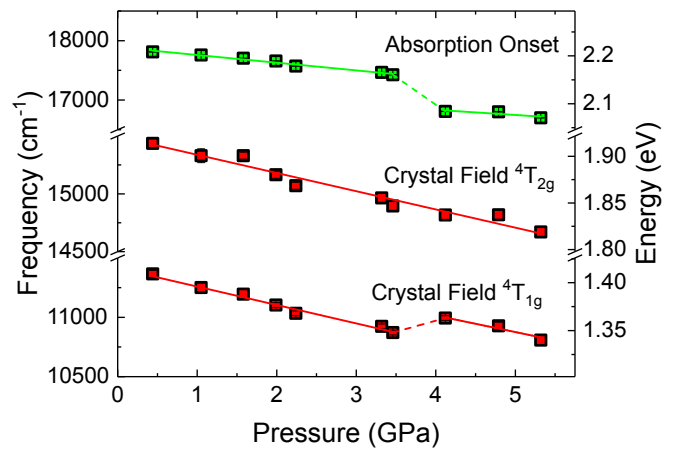}
\caption{Position of crystal field excitations ($^{6}\mathrm{A}_{1\mathrm{g}}\rightarrow{}^{4}\mathrm{T}_{1\mathrm{g}}$,
$^{6}\mathrm{A}_{1\mathrm{g}}\rightarrow{}^{4}\mathrm{T}_{2\mathrm{g}}$) and the absorption onset as a function of pressure with linear fits as guides to the eye.}
\label{fig:crystal_field}
\end{figure}

The pressure-induced changes in the absorbance spectra are related to the strong influence of external pressure on the ambient-pressure $R3c$ crystal structure: the rhomboedral lattice parameter $a_{\mathrm{rh}}$ decreases with increasing pressure, whereas the rhomboedral cell distortion angle $\alpha{}_{\mathrm{rh}}$ shows an increase.\cite{Guennou.2011(b)} In contrast, the FeO$_6$ tilting angle decreases under pressure and the value of the Fe-O bond length decreases as well.\cite{Haumont.2009} Since the crystal field transitions depend mainly on the FeO$_6$ local structure, they are highly sensitive to changes of the Fe$^{3+}$ coordination.\cite{GomezSalces.2012} The pressure-induced decrease of the Fe-O bond length leads to an increase of the $e_\mathrm{g}$-$t_{2\mathrm{g}}$ crystal-field splitting.\cite{GomezSalces.2012} According to the Tanabe-Sugano diagram [see Fig.\ 5(a) in Ref.\ \citenum{GomezSalces.2012}] the $^{4}\mathrm{T}_{1\mathrm{g}}$ and $^{4}\mathrm{T}_{2\mathrm{g}}$ crystal field transitions are expected to shift to lower energies under pressure.\cite{GomezSalces.2012}

\begin{figure*}
\includegraphics[width=0.95\textwidth]{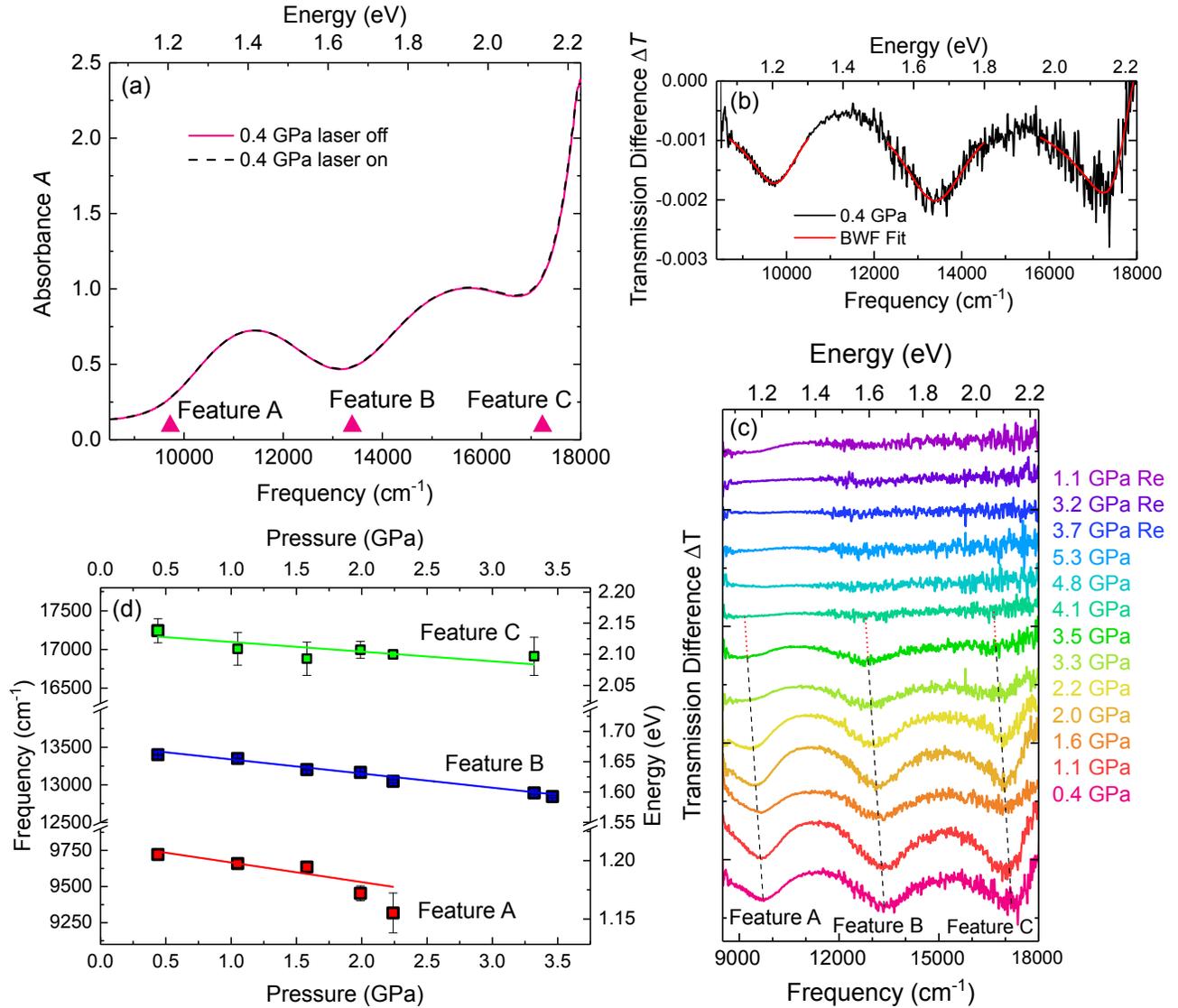}
\caption{(a) Absorbance spectrum of BFO at 0.4~GPa with and without laser illumination. The red triangles indicate the positions of feature A, B, and C. (b) BWF fit of the three absorption features of the transmission difference spectrum $\Delta{}T$ at 0.4~GPa. (c) Transmission difference spectra $\Delta{}T$ showing the light-induced absorption features A, B, and C for pressures between 0.4 and 5.3~GPa. The spectra recorded during pressure releasing are labeled with ``Re''. The dashed lines illustrate the pressure-induced shifts of the absorption features. (d) Extracted feature positions as a function of pressure with linear fits as guides to the eye.}
\label{fig:photostriction_spectrum_all}
\end{figure*}

The overall absorbance increase between 3.5 and 4.1~GPa might be due to a change in the electronic structure or due to the structural phase transition at $P_{\mathrm{c1}}$.\cite{Guennou.2011, Haumont.2009}
Also the energy position of the absorption onset changes significantly under pressure, since it is highly sensitive to structural changes.\cite{GomezSalces.2012} For a quantitative analysis of the pressure-induced changes regarding the crystal field excitations and the absorption onset, we fitted the absorbance spectra measured without laser illumination with two Gaussian functions for describing the crystal field excitations and one Lorentzian term for the absorption onset [see Fig.\ \ref{fig:absorbance}(b)], similar to Ref.\ \citenum{GomezSalces.2012}.
The parameters of the Lorentzian function have a rather high uncertainty, since we can fit the onset only up to 18~000~cm$^{-1}$ ($\approx{}2.23$~eV). Therefore, instead of using the energy position of the Lorentzian function as a measure for the position of the absorption onset, we used the frequency where the absorbance level reaches the value $A$=2.0. We consider this criterion as reliable, since the frequency, where $A$=2.0, is high enough not to get disturbed by the crystal field excitation $^4$A$_{1\mathrm{g}}\rightarrow{} ^4$T$_{2\mathrm{g}}$ and low enough not to be masked by noise close to the high-frequency limit of our measurements. Furthermore, we note that in the pressure regime $P\leq{}$3.5~GPa, which is relevant for the observed features under laser illumination (as the features disappear above 3.5~GPa), the two different analysis methods for the absorption onset (Lorentz position \textit{versus} frequency of the $A$=2.0 level) only differ by a {\it pressure-independent} offset. Thus, the choice of the a\-na\-ly\-sis method will not change the main conclusions drawn in the following.

The pressure-dependent energy positions of the crystal field excitations and the absorption onset are plotted in Fig.\ \ref{fig:crystal_field}. Up to $P_{\mathrm{c1}}\approx{}$3.5~GPa, all three intrinsic excitations shift monotonically to lower energy with increasing pressure. At $P_{\mathrm{c1}}$ the pressure-induced redshift of the absorption onset and the $^4$T$_{1\mathrm{g}}$ excitation shows an anomaly, whereas the monotonic redshift of the $^4$T$_{2\mathrm{g}}$ excitation is barely affected, consistent with earlier reports.\cite{GomezSalces.2012}

In the following, we focus on the laser-induced features in the absorbance spectrum of BFO.
Recent optical measurements at ambient pressure\cite{Burkert.2016, Meggle.2019} observed three absorption features during laser illumination. Since the spectral changes caused by the laser illumination were very small, the transmission difference spectrum $\Delta{}T(\nu)$ was considered:
\begin{equation}
\Delta T(\nu) = [I_{\mathrm{BFO,on}}(\nu) - I_{\mathrm{BFO,off}}(\nu)]/I_{\mathrm{ref}}(\nu).
\label{eq:Delta_trans_norm}
\end{equation}
Hereby, $I_{\mathrm{BFO,on/off}}(\nu)$ is the intensity transmitted by the BFO crystal without laser illumination (``off'') or during laser illumination (``on''), respectively, and  $I_{\mathrm{ref}}(\nu)$ represents the intensity of the reference. The light-induced features were previously\cite{Burkert.2016, Meggle.2019} interpreted in terms of excitons, which are possible related to the ultra-fast photostriction effect in BFO.\cite{Schick.2014}

Figure\ \ref{fig:photostriction_spectrum_all}(a) shows the absorption spectrum of illuminated and non-illuminated BFO at 0.4~GPa as an example. The light-induced spectral changes are extremely small, consistent with Refs.\ \cite{Burkert.2016, Meggle.2019}, so we consider the transmission difference spectra $\Delta{}T$, which is depicted with a vertical offset in Fig.\ \ref{fig:photostriction_spectrum_all}(c) for pressures up to 5.3~GPa. Between 0.4 and 3.5~GPa the transmission difference spectra consist of three asymmetric absorption features, which are labeled as feature A, B, and C, respectively.  With increasing pressure the features shift to lower energies [indicated by dashed lines in Fig.\ \ref{fig:photostriction_spectrum_all}(c)] and lose intensity. At 3.5~GPa feature A and B are still clearly observable, and feature C is close to disappear but is still slightly visible. For pressures above 3.5~GPa all features have disappeared, i.e., the transmission spectra with and without laser illumination are equal to each other. During pressure release only one broad dip located at around 9000~cm$^{-1}$ appears at the lowest pressure (1.1 GPa Re), i.e., the pressure-induced suppression of the absorption features is irreversible upon pressure release.

In order to determine the exact position of the features, we apply the same analysis of the data as described in Ref.\ \citenum{Meggle.2019}: We fit the features A, B and C by using a Breit-Wigner-Fano (BWF) line shape each [see Fig.\ \ref{fig:photostriction_spectrum_all}(b)] and determined the feature positions by equalizing the derived BWF formula to zero and insert the parameters from the fit. For feature A a good fit can only be obtained for pressures below 3.3~GPa and for feature C below 3.5~GPa.

At the lowest pressure (0.4~GPa) the features are located at $\nu{}_{\mathrm{A}}=9721$~cm$^{-1}$ ($E_{\mathrm{A}}=1.21$~eV), $\nu{}_{\mathrm{B}}=13400$~cm$^{-1}$ ($E_{\mathrm{B}}=1.66$~eV) and $\nu{}_{\mathrm{C}}=17243$~cm$^{-1}$ ($E_{\mathrm{C}}=2.14$~eV). This is in good agreement with earlier measurements at ambient conditions.\cite{Burkert.2016}
The feature positions at 0.4~GPa are indicated by red triangles in Fig.\ \ref{fig:photostriction_spectrum_all}(a). Obviously, they lie on the low-energy side of the crystal field excitation/absorption onset, respectively. The energy positions of features A, B, and C as obtained from the fitting are plotted in Fig.\ \ref{fig:photostriction_spectrum_all}(d) as a function of pressure. All three features exhibit an individual redshift in the pressure range between 0.4 and 3.5~GPa.

\begin{figure}[t]
\includegraphics[width=1\columnwidth]{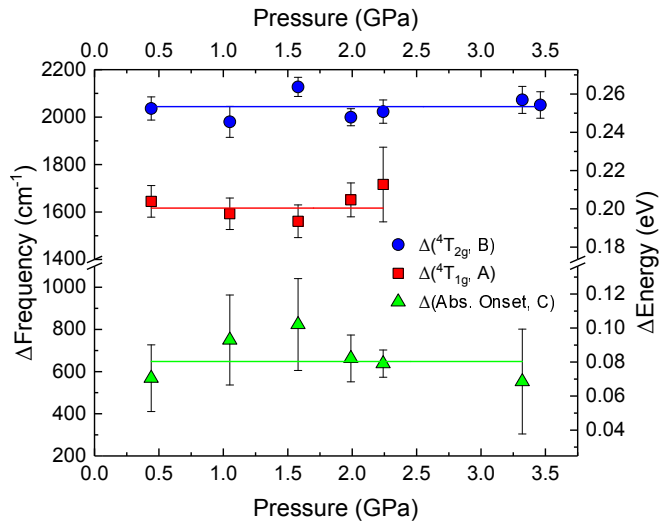}
\caption{Energy difference $\Delta{}(X,Y)$ between the crystal field excitation/absorption onset $X$ and the corresponding absorption feature $Y$. As guides to the eyes, constant functions were used.}
\label{fig:differeneces}
\end{figure}

According to the electronic band scheme suggested in Ref.\ \citenum{Meggle.2019} the excitonic excitations are linked to the intrinsic excitations in BFO. Indeed, the pressure-dependent energies of the three absorption features follow the pressure dependence of the crystal field excitations and the absorption onset, respectively, as illustrated in Fig.\ \ref{fig:differeneces}: The energy difference $\Delta{}$ between the crystal field excitation/absorption onset and the corresponding absorption feature does not show a clear pressure dependence within the error bar.\footnote{As mentioned above, the energy of the absorption onset could also be estimated from the energy position of the Lorentzian used in the fitting of the absorbance spectrum. Using this energy position for calculating the energy difference $\Delta$ gives a similar pressure dependence as shown in Fig.\ \ref{fig:differeneces}, but with a constant offset.} Accordingly, the pressure-dependent shifts of the laser-induced absorption features are mainly determined by the pressure-dependent shifts of the intrinsic excitations in BFO.

\begin{figure}[t]
\includegraphics[width=1\columnwidth]{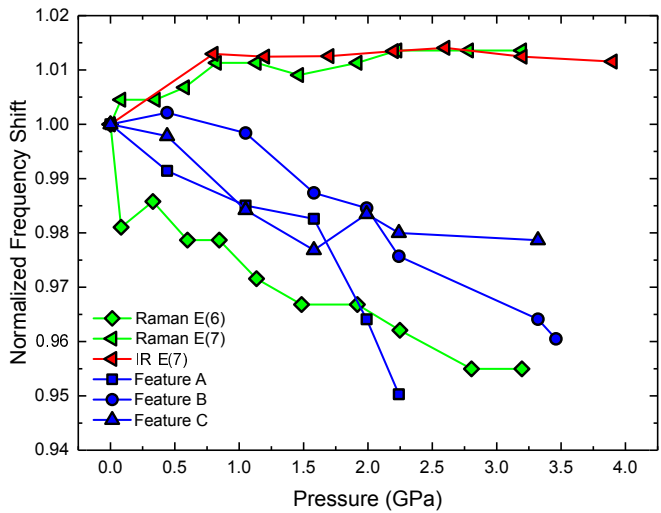}
\caption{Pressure evolution of the frequency positions of the light-induced absorption features (blue symbols) and selected phonon modes. The positions of the absorption features were normalized to their ambient-pressure values taken from Ref. \citenum{Burkert.2016}. Also the frequencies of the IR- and Raman-active modes were normalized to their ambient-pressure values. The phonon modes from Raman measurements (green symbols) were extracted from Ref.\ \citenum{GomezSalces.2012}, and for comparison we also show the phonon mode E(7) from IR measurements \cite{Haumont.2009} (red symbols). The ambient-pressure position of E(7) (IR measurement) was extracted from Lobo et al.\ \cite{Lobo.2007}.}
\label{fig:phonons}
\end{figure}

As a consequence, a possible pressure dependence of the excitonic features due to their coupling to phonon modes, as suggested previously based on the temperature-dependent behavior,\cite{Meggle.2019} is masked by the rather strong pressure dependence of the intrinsic excitations in BFO (please note that the pressure dependence of the intrinsic excitations is much stronger than their temperature dependence). In particular, most of the infrared (IR)- and Raman-active phonon modes in BFO show a hardening with increasing pressure below $P_{\mathrm{c1}}$. As an example, we show in Fig.\ \ref{fig:phonons} the frequency of the phonon mode E(7) as a function of pressure normalized to its ambient-pressure frequency, as obtained by Raman and IR measurements.\cite{Haumont.2009, GomezSalces.2012} Only the phonon mode E(6) softens gradually under compression (see Fig.\ \ref{fig:phonons} for the pressure-dependent frequency position normalized to its ambient-pressure value) and could possibly explain the observed pressure-induced redshift of the laser-induced absorption features.

Remarkably, the intensity of the light-induced absorption features are strongly affected by the pressure application: with increasing pressure the intensity of the absorption features decreases gradually, and the features disappear at the critical pressure $P_{\mathrm{c1}}$ of the structural phase transition. Interestingly, also the electric polarization of BFO connected to the $R3c$ phase decreases with increasing pressure\cite{Kozlenko.2011} and disappears above $P_{\mathrm{c1}}$.\cite{Haumont.2009, Chen.2012(b), Guennou.2011(b), Mishra.2013, Belik.2009, Kozlenko.2011, Wu.2018, Zhang.2013, Guo.2019} Additional studies are needed to elucidate the atomistic and electronic origin of the disappearance of the features above $P_{\mathrm{c1}}$.

During pressure release only feature A reappears at 1.1~GPa [see Fig.\ \ref{fig:photostriction_spectrum_all}(c)], indicating an irreversible process. Literature is not consistent regarding the reversibility of the pressure-induced structural changes in BFO. Haumont et al.\cite{Haumont.2009} observed the full reversibility of their XRD pattern after reaching 37~GPa. In contrast, high-pressure XRD measurements on BFO from Belik et al.\cite{Belik.2009} showed a mixture of $Pbam$ and $R3c$ phases below 0.9~GPa during pressure-release. The pressure value for the appearance of the $R3c$+$Pbam$ mixture is in fair agreement with the pressure 1.1~GPa, where we observe hints for the reappearance of the absorption feature A.
%Furthermore, it is rather interesting that only feature A appears during releasing and feature C disappears first during compression. This might be a hint %that each absorption feature could be a stand-alone feature.

\section{Conclusion}
We studied the optical transmission spectrum of an illuminated BiFeO$_3$ single crystal for hydrostatic pressure between 0.4 and 5.3~GPa. At low pressures, we observe three light-induced absorption features, which were previously ascribed to excitons. With increasing pressure all three absorption features shift to lower energies, following the pressure dependence of the corresponding crystal field excitation or absorption onset. The intensity of the three features decreases with increasing pressure and they are no longer visible above the critical pressure $P_{\mathrm{c1}}\approx$3.5~GPa of the structural phase transition, suggesting a link between the light-induced absorption features and the ferroelectric $R3c$ phase.

\begin{acknowledgments}
We acknowledge fruitful discussions with M. Viret.
This work is financially supported by the Deutsche Forschungsgemeinschaft (DFG) through grant no.\ KU 1432/9-1. Jens Kreisel acknowledges support from the National Research Fund Luxembourg through a Pearl Grant (FNR/P12/4853155).
\end{acknowledgments}


\begin{thebibliography}{55}%
\makeatletter
\providecommand \@ifxundefined [1]{%
 \@ifx{#1\undefined}
}%
\providecommand \@ifnum [1]{%
 \ifnum #1\expandafter \@firstoftwo
 \else \expandafter \@secondoftwo
 \fi
}%
\providecommand \@ifx [1]{%
 \ifx #1\expandafter \@firstoftwo
 \else \expandafter \@secondoftwo
 \fi
}%
\providecommand \natexlab [1]{#1}%
\providecommand \enquote  [1]{``#1''}%
\providecommand \bibnamefont  [1]{#1}%
\providecommand \bibfnamefont [1]{#1}%
\providecommand \citenamefont [1]{#1}%
\providecommand \href@noop [0]{\@secondoftwo}%
\providecommand \href [0]{\begingroup \@sanitize@url \@href}%
\providecommand \@href[1]{\@@startlink{#1}\@@href}%
\providecommand \@@href[1]{\endgroup#1\@@endlink}%
\providecommand \@sanitize@url [0]{\catcode `\\12\catcode `\$12\catcode
  `\&12\catcode `\#12\catcode `\^12\catcode `\_12\catcode `\%12\relax}%
\providecommand \@@startlink[1]{}%
\providecommand \@@endlink[0]{}%
\providecommand \url  [0]{\begingroup\@sanitize@url \@url }%
\providecommand \@url [1]{\endgroup\@href {#1}{\urlprefix }}%
\providecommand \urlprefix  [0]{URL }%
\providecommand \Eprint [0]{\href }%
\providecommand \doibase [0]{http://dx.doi.org/}%
\providecommand \selectlanguage [0]{\@gobble}%
\providecommand \bibinfo  [0]{\@secondoftwo}%
\providecommand \bibfield  [0]{\@secondoftwo}%
\providecommand \translation [1]{[#1]}%
\providecommand \BibitemOpen [0]{}%
\providecommand \bibitemStop [0]{}%
\providecommand \bibitemNoStop [0]{.\EOS\space}%
\providecommand \EOS [0]{\spacefactor3000\relax}%
\providecommand \BibitemShut  [1]{\csname bibitem#1\endcsname}%
\let\auto@bib@innerbib\@empty
%</preamble>
\bibitem [{\citenamefont {Mitchell}(2002)}]{Mitchell.2002}%
  \BibitemOpen
  \bibfield  {author} {\bibinfo {author} {\bibfnamefont {R.~H.}\ \bibnamefont
  {Mitchell}},\ }\href
  {https://www.amazon.com/Perovskites-Ancient-Roger-H-Mitchell/dp/0968941109?SubscriptionId=AKIAIOBINVZYXZQZ2U3A&tag=chimbori05-20&linkCode=xm2&camp=2025&creative=165953&creativeASIN=0968941109}
  {\emph {\bibinfo {title} {Perovskites: Modern and Ancient}}}\ (\bibinfo
  {publisher} {Almaz Press},\ \bibinfo {year} {2002})\BibitemShut {NoStop}%
\bibitem [{\citenamefont {Li}\ \emph {et~al.}(2015)\citenamefont {Li},
  \citenamefont {Wang}, \citenamefont {Jin}, \citenamefont {Lin}, \citenamefont
  {Li}, \citenamefont {Li}, \citenamefont {Xu},\ and\ \citenamefont
  {Zhang}}]{Li.2015(a)}%
  \BibitemOpen
  \bibfield  {author} {\bibinfo {author} {\bibfnamefont {F.}~\bibnamefont
  {Li}}, \bibinfo {author} {\bibfnamefont {L.}~\bibnamefont {Wang}}, \bibinfo
  {author} {\bibfnamefont {L.}~\bibnamefont {Jin}}, \bibinfo {author}
  {\bibfnamefont {D.}~\bibnamefont {Lin}}, \bibinfo {author} {\bibfnamefont
  {J.}~\bibnamefont {Li}}, \bibinfo {author} {\bibfnamefont {Z.}~\bibnamefont
  {Li}}, \bibinfo {author} {\bibfnamefont {Z.}~\bibnamefont {Xu}}, \ and\
  \bibinfo {author} {\bibfnamefont {S.}~\bibnamefont {Zhang}},\ }\href
  {\doibase 10.1109/TUFFC.2014.006660} {\bibfield  {journal} {\bibinfo
  {journal} {IEEE Trans. Ultrason. Ferroelectr. Freq. Control}\ }\textbf
  {\bibinfo {volume} {62}},\ \bibinfo {pages} {18} (\bibinfo {year}
  {2015})}\BibitemShut {NoStop}%
\bibitem [{\citenamefont {Liu}\ and\ \citenamefont {Yang}(2017)}]{Liu.2017}%
  \BibitemOpen
  \bibfield  {author} {\bibinfo {author} {\bibfnamefont {H.}~\bibnamefont
  {Liu}}\ and\ \bibinfo {author} {\bibfnamefont {X.}~\bibnamefont {Yang}},\
  }\href {\doibase 10.1080/00150193.2017.1283171} {\bibfield  {journal}
  {\bibinfo  {journal} {Ferroelectrics}\ }\textbf {\bibinfo {volume} {507}},\
  \bibinfo {pages} {69} (\bibinfo {year} {2017})}\BibitemShut {NoStop}%
\bibitem [{\citenamefont {Hill}(2000)}]{Hill.2000}%
  \BibitemOpen
  \bibfield  {author} {\bibinfo {author} {\bibfnamefont {N.~A.}\ \bibnamefont
  {Hill}},\ }\href {\doibase 10.1021/jp000114x} {\bibfield  {journal} {\bibinfo
   {journal} {J. Phys. Chem. B}\ }\textbf {\bibinfo {volume} {104}},\ \bibinfo
  {pages} {6694} (\bibinfo {year} {2000})}\BibitemShut {NoStop}%
\bibitem [{\citenamefont {Kubel}\ and\ \citenamefont
  {Schmid}(1990)}]{Kubel.1990}%
  \BibitemOpen
  \bibfield  {author} {\bibinfo {author} {\bibfnamefont {F.}~\bibnamefont
  {Kubel}}\ and\ \bibinfo {author} {\bibfnamefont {H.}~\bibnamefont {Schmid}},\
  }\href {\doibase 10.1107/S0108768190006887} {\bibfield  {journal} {\bibinfo
  {journal} {Acta Cryst. B}\ }\textbf {\bibinfo {volume} {46}},\ \bibinfo
  {pages} {698} (\bibinfo {year} {1990})}\BibitemShut {NoStop}%
\bibitem [{\citenamefont {Glazer}(1972)}]{Glazer.1972}%
  \BibitemOpen
  \bibfield  {author} {\bibinfo {author} {\bibfnamefont {A.~M.}\ \bibnamefont
  {Glazer}},\ }\href {\doibase 10.1107/S0567740872007976} {\bibfield  {journal}
  {\bibinfo  {journal} {Acta Cryst. B}\ }\textbf {\bibinfo {volume} {28}},\
  \bibinfo {pages} {3384} (\bibinfo {year} {1972})}\BibitemShut {NoStop}%
\bibitem [{\citenamefont {{\'{I}}{\~{n}}iguez}, \citenamefont {Vanderbilt},\
  and\ \citenamefont {Bellaiche}(2003)}]{Iniguez.2003}%
  \BibitemOpen
  \bibfield  {author} {\bibinfo {author} {\bibfnamefont {J.}~\bibnamefont
  {{\'{I}}{\~{n}}iguez}}, \bibinfo {author} {\bibfnamefont {D.}~\bibnamefont
  {Vanderbilt}}, \ and\ \bibinfo {author} {\bibfnamefont {L.}~\bibnamefont
  {Bellaiche}},\ }\href {\doibase 10.1103/physrevb.67.224107} {\bibfield
  {journal} {\bibinfo  {journal} {Phys. Rev. B}\ }\textbf {\bibinfo {volume}
  {67}},\ \bibinfo {pages} {224107} (\bibinfo {year} {2003})}\BibitemShut
  {NoStop}%
\bibitem [{\citenamefont {Seshadri}\ and\ \citenamefont
  {Hill}(2001)}]{Seshadri.2001}%
  \BibitemOpen
  \bibfield  {author} {\bibinfo {author} {\bibfnamefont {R.}~\bibnamefont
  {Seshadri}}\ and\ \bibinfo {author} {\bibfnamefont {N.~A.}\ \bibnamefont
  {Hill}},\ }\href {\doibase 10.1021/cm010090m} {\bibfield  {journal} {\bibinfo
   {journal} {Chem. Mater.}\ }\textbf {\bibinfo {volume} {13}},\ \bibinfo
  {pages} {2892} (\bibinfo {year} {2001})}\BibitemShut {NoStop}%
\bibitem [{\citenamefont {Neaton}\ \emph {et~al.}(2005)\citenamefont {Neaton},
  \citenamefont {Ederer}, \citenamefont {Waghmare}, \citenamefont {Spaldin},\
  and\ \citenamefont {Rabe}}]{Neaton.2005}%
  \BibitemOpen
  \bibfield  {author} {\bibinfo {author} {\bibfnamefont {J.~B.}\ \bibnamefont
  {Neaton}}, \bibinfo {author} {\bibfnamefont {C.}~\bibnamefont {Ederer}},
  \bibinfo {author} {\bibfnamefont {U.~V.}\ \bibnamefont {Waghmare}}, \bibinfo
  {author} {\bibfnamefont {N.~A.}\ \bibnamefont {Spaldin}}, \ and\ \bibinfo
  {author} {\bibfnamefont {K.~M.}\ \bibnamefont {Rabe}},\ }\href {\doibase
  10.1103/PhysRevB.71.014113} {\bibfield  {journal} {\bibinfo  {journal} {Phys.
  Rev. B}\ }\textbf {\bibinfo {volume} {71}},\ \bibinfo {pages} {014113}
  (\bibinfo {year} {2005})}\BibitemShut {NoStop}%
\bibitem [{\citenamefont {Shvartsman}\ \emph {et~al.}(2007)\citenamefont
  {Shvartsman}, \citenamefont {Kleemann}, \citenamefont {Haumont},\ and\
  \citenamefont {Kreisel}}]{Shvartsman.2007}%
  \BibitemOpen
  \bibfield  {author} {\bibinfo {author} {\bibfnamefont {V.~V.}\ \bibnamefont
  {Shvartsman}}, \bibinfo {author} {\bibfnamefont {W.}~\bibnamefont
  {Kleemann}}, \bibinfo {author} {\bibfnamefont {R.}~\bibnamefont {Haumont}}, \
  and\ \bibinfo {author} {\bibfnamefont {J.}~\bibnamefont {Kreisel}},\ }\href
  {\doibase 10.1063/1.2731312} {\bibfield  {journal} {\bibinfo  {journal}
  {Appl. Phys. Lett.}\ }\textbf {\bibinfo {volume} {90}},\ \bibinfo {pages}
  {172115} (\bibinfo {year} {2007})}\BibitemShut {NoStop}%
\bibitem [{\citenamefont {Lebeugle}\ \emph {et~al.}(2007)\citenamefont
  {Lebeugle}, \citenamefont {Colson}, \citenamefont {Forget},\ and\
  \citenamefont {Viret}}]{Lebeugle.2007(b)}%
  \BibitemOpen
  \bibfield  {author} {\bibinfo {author} {\bibfnamefont {D.}~\bibnamefont
  {Lebeugle}}, \bibinfo {author} {\bibfnamefont {D.}~\bibnamefont {Colson}},
  \bibinfo {author} {\bibfnamefont {A.}~\bibnamefont {Forget}}, \ and\ \bibinfo
  {author} {\bibfnamefont {M.}~\bibnamefont {Viret}},\ }\href {\doibase
  10.1063/1.2753390} {\bibfield  {journal} {\bibinfo  {journal} {Appl. Phys.
  Lett.}\ }\textbf {\bibinfo {volume} {91}},\ \bibinfo {pages} {022907}
  (\bibinfo {year} {2007})}\BibitemShut {NoStop}%
\bibitem [{\citenamefont {Cohen}\ and\ \citenamefont
  {Krakauer}(1990)}]{Cohen.1990}%
  \BibitemOpen
  \bibfield  {author} {\bibinfo {author} {\bibfnamefont {R.~E.}\ \bibnamefont
  {Cohen}}\ and\ \bibinfo {author} {\bibfnamefont {H.}~\bibnamefont
  {Krakauer}},\ }\href {\doibase 10.1103/physrevb.42.6416} {\bibfield
  {journal} {\bibinfo  {journal} {Phys. Rev. B}\ }\textbf {\bibinfo {volume}
  {42}},\ \bibinfo {pages} {6416} (\bibinfo {year} {1990})}\BibitemShut
  {NoStop}%
\bibitem [{\citenamefont {Cohen}(1992)}]{Cohen.1992}%
  \BibitemOpen
  \bibfield  {author} {\bibinfo {author} {\bibfnamefont {R.~E.}\ \bibnamefont
  {Cohen}},\ }\href {\doibase 10.1038/358136a0} {\bibfield  {journal} {\bibinfo
   {journal} {Nature}\ }\textbf {\bibinfo {volume} {358}},\ \bibinfo {pages}
  {136} (\bibinfo {year} {1992})}\BibitemShut {NoStop}%
\bibitem [{\citenamefont {Samara}, \citenamefont {Sakudo},\ and\ \citenamefont
  {Yoshimitsu}(1975)}]{Samara.1975}%
  \BibitemOpen
  \bibfield  {author} {\bibinfo {author} {\bibfnamefont {G.~A.}\ \bibnamefont
  {Samara}}, \bibinfo {author} {\bibfnamefont {T.}~\bibnamefont {Sakudo}}, \
  and\ \bibinfo {author} {\bibfnamefont {K.}~\bibnamefont {Yoshimitsu}},\
  }\href {\doibase 10.1103/physrevlett.35.1767} {\bibfield  {journal} {\bibinfo
   {journal} {Phys. Rev. Lett.}\ }\textbf {\bibinfo {volume} {35}},\ \bibinfo
  {pages} {1767} (\bibinfo {year} {1975})}\BibitemShut {NoStop}%
\bibitem [{\citenamefont {Kornev}\ \emph {et~al.}(2005)\citenamefont {Kornev},
  \citenamefont {Bellaiche}, \citenamefont {Bouvier}, \citenamefont {Janolin},
  \citenamefont {Dkhil},\ and\ \citenamefont {Kreisel}}]{Kornev.2005}%
  \BibitemOpen
  \bibfield  {author} {\bibinfo {author} {\bibfnamefont {I.~A.}\ \bibnamefont
  {Kornev}}, \bibinfo {author} {\bibfnamefont {L.}~\bibnamefont {Bellaiche}},
  \bibinfo {author} {\bibfnamefont {P.}~\bibnamefont {Bouvier}}, \bibinfo
  {author} {\bibfnamefont {P.-E.}\ \bibnamefont {Janolin}}, \bibinfo {author}
  {\bibfnamefont {B.}~\bibnamefont {Dkhil}}, \ and\ \bibinfo {author}
  {\bibfnamefont {J.}~\bibnamefont {Kreisel}},\ }\href {\doibase
  10.1103/PhysRevLett.95.196804} {\bibfield  {journal} {\bibinfo  {journal}
  {Phys. Rev. Lett.}\ }\textbf {\bibinfo {volume} {95}},\ \bibinfo {pages}
  {196804} (\bibinfo {year} {2005})}\BibitemShut {NoStop}%
\bibitem [{\citenamefont {Janolin}\ \emph {et~al.}(2008)\citenamefont
  {Janolin}, \citenamefont {Bouvier}, \citenamefont {Kreisel}, \citenamefont
  {Thomas}, \citenamefont {Kornev}, \citenamefont {Bellaiche}, \citenamefont
  {Crichton}, \citenamefont {Hanfland},\ and\ \citenamefont
  {Dkhil}}]{Janolin.2008}%
  \BibitemOpen
  \bibfield  {author} {\bibinfo {author} {\bibfnamefont {P.-E.}\ \bibnamefont
  {Janolin}}, \bibinfo {author} {\bibfnamefont {P.}~\bibnamefont {Bouvier}},
  \bibinfo {author} {\bibfnamefont {J.}~\bibnamefont {Kreisel}}, \bibinfo
  {author} {\bibfnamefont {P.~A.}\ \bibnamefont {Thomas}}, \bibinfo {author}
  {\bibfnamefont {I.~A.}\ \bibnamefont {Kornev}}, \bibinfo {author}
  {\bibfnamefont {L.}~\bibnamefont {Bellaiche}}, \bibinfo {author}
  {\bibfnamefont {W.}~\bibnamefont {Crichton}}, \bibinfo {author}
  {\bibfnamefont {M.}~\bibnamefont {Hanfland}}, \ and\ \bibinfo {author}
  {\bibfnamefont {B.}~\bibnamefont {Dkhil}},\ }\href {\doibase
  10.1103/physrevlett.101.237601} {\bibfield  {journal} {\bibinfo  {journal}
  {Phys. Rev. Lett.}\ }\textbf {\bibinfo {volume} {101}},\ \bibinfo {pages}
  {237601} (\bibinfo {year} {2008})}\BibitemShut {NoStop}%
\bibitem [{\citenamefont {Guennou}\ \emph
  {et~al.}(2011{\natexlab{a}})\citenamefont {Guennou}, \citenamefont {Bouvier},
  \citenamefont {Chen}, \citenamefont {Dkhil}, \citenamefont {Haumont},
  \citenamefont {Garbarino},\ and\ \citenamefont {Kreisel}}]{Guennou.2011}%
  \BibitemOpen
  \bibfield  {author} {\bibinfo {author} {\bibfnamefont {M.}~\bibnamefont
  {Guennou}}, \bibinfo {author} {\bibfnamefont {P.}~\bibnamefont {Bouvier}},
  \bibinfo {author} {\bibfnamefont {G.~S.}\ \bibnamefont {Chen}}, \bibinfo
  {author} {\bibfnamefont {B.}~\bibnamefont {Dkhil}}, \bibinfo {author}
  {\bibfnamefont {R.}~\bibnamefont {Haumont}}, \bibinfo {author} {\bibfnamefont
  {G.}~\bibnamefont {Garbarino}}, \ and\ \bibinfo {author} {\bibfnamefont
  {J.}~\bibnamefont {Kreisel}},\ }\href {\doibase 10.1103/PhysRevB.84.174107}
  {\bibfield  {journal} {\bibinfo  {journal} {Phys. Rev. B}\ }\textbf {\bibinfo
  {volume} {84}},\ \bibinfo {pages} {174107} (\bibinfo {year}
  {2011}{\natexlab{a}})}\BibitemShut {NoStop}%
\bibitem [{\citenamefont {G{\'o}mez-Salces}\ \emph {et~al.}(2012)\citenamefont
  {G{\'o}mez-Salces}, \citenamefont {Aguado}, \citenamefont {Rodr{\'i}guez},
  \citenamefont {Valiente}, \citenamefont {Gonz{\'a}lez}, \citenamefont
  {Haumont},\ and\ \citenamefont {Kreisel}}]{GomezSalces.2012}%
  \BibitemOpen
  \bibfield  {author} {\bibinfo {author} {\bibfnamefont {S.}~\bibnamefont
  {G{\'o}mez-Salces}}, \bibinfo {author} {\bibfnamefont {F.}~\bibnamefont
  {Aguado}}, \bibinfo {author} {\bibfnamefont {F.}~\bibnamefont
  {Rodr{\'i}guez}}, \bibinfo {author} {\bibfnamefont {R.}~\bibnamefont
  {Valiente}}, \bibinfo {author} {\bibfnamefont {J.}~\bibnamefont
  {Gonz{\'a}lez}}, \bibinfo {author} {\bibfnamefont {R.}~\bibnamefont
  {Haumont}}, \ and\ \bibinfo {author} {\bibfnamefont {J.}~\bibnamefont
  {Kreisel}},\ }\href {\doibase 10.1103/PhysRevB.85.144109} {\bibfield
  {journal} {\bibinfo  {journal} {Phys. Rev. B}\ }\textbf {\bibinfo {volume}
  {85}},\ \bibinfo {pages} {144109} (\bibinfo {year} {2012})}\BibitemShut
  {NoStop}%
\bibitem [{\citenamefont {Wu}, \citenamefont {Han},\ and\ \citenamefont
  {Huang}(2018)}]{Wu.2018}%
  \BibitemOpen
  \bibfield  {author} {\bibinfo {author} {\bibfnamefont {Y.}~\bibnamefont
  {Wu}}, \bibinfo {author} {\bibfnamefont {X.}~\bibnamefont {Han}}, \ and\
  \bibinfo {author} {\bibfnamefont {H.}~\bibnamefont {Huang}},\ }\href
  {\doibase 10.1021/acs.jpcc.8b00977} {\bibfield  {journal} {\bibinfo
  {journal} {J. Phys. Chem. C}\ }\textbf {\bibinfo {volume} {122}},\ \bibinfo
  {pages} {6852} (\bibinfo {year} {2018})}\BibitemShut {NoStop}%
\bibitem [{\citenamefont {Mishra}\ \emph {et~al.}(2013)\citenamefont {Mishra},
  \citenamefont {Shanavas}, \citenamefont {Poswal}, \citenamefont {Mandal},
  \citenamefont {Garg},\ and\ \citenamefont {Sharma}}]{Mishra.2013}%
  \BibitemOpen
  \bibfield  {author} {\bibinfo {author} {\bibfnamefont {A.}~\bibnamefont
  {Mishra}}, \bibinfo {author} {\bibfnamefont {K.}~\bibnamefont {Shanavas}},
  \bibinfo {author} {\bibfnamefont {H.}~\bibnamefont {Poswal}}, \bibinfo
  {author} {\bibfnamefont {B.}~\bibnamefont {Mandal}}, \bibinfo {author}
  {\bibfnamefont {N.}~\bibnamefont {Garg}}, \ and\ \bibinfo {author}
  {\bibfnamefont {S.~M.}\ \bibnamefont {Sharma}},\ }\href {\doibase
  10.1016/j.ssc.2012.10.018} {\bibfield  {journal} {\bibinfo  {journal} {Solid
  State Commun.}\ }\textbf {\bibinfo {volume} {154}},\ \bibinfo {pages} {72}
  (\bibinfo {year} {2013})}\BibitemShut {NoStop}%
\bibitem [{\citenamefont {Belik}\ \emph {et~al.}(2009)\citenamefont {Belik},
  \citenamefont {Yusa}, \citenamefont {Hirao}, \citenamefont {Ohishi},\ and\
  \citenamefont {Takayama-Muromachi}}]{Belik.2009}%
  \BibitemOpen
  \bibfield  {author} {\bibinfo {author} {\bibfnamefont {A.~A.}\ \bibnamefont
  {Belik}}, \bibinfo {author} {\bibfnamefont {H.}~\bibnamefont {Yusa}},
  \bibinfo {author} {\bibfnamefont {N.}~\bibnamefont {Hirao}}, \bibinfo
  {author} {\bibfnamefont {Y.}~\bibnamefont {Ohishi}}, \ and\ \bibinfo {author}
  {\bibfnamefont {E.}~\bibnamefont {Takayama-Muromachi}},\ }\href {\doibase
  10.1021/cm901008t} {\bibfield  {journal} {\bibinfo  {journal} {Chem. Mater.}\
  }\textbf {\bibinfo {volume} {21}},\ \bibinfo {pages} {3400} (\bibinfo {year}
  {2009})}\BibitemShut {NoStop}%
\bibitem [{\citenamefont {Kozlenko}\ \emph {et~al.}(2011)\citenamefont
  {Kozlenko}, \citenamefont {Belik}, \citenamefont {Belushkin}, \citenamefont
  {Lukin}, \citenamefont {Marshall}, \citenamefont {Savenko},\ and\
  \citenamefont {Takayama-Muromachi}}]{Kozlenko.2011}%
  \BibitemOpen
  \bibfield  {author} {\bibinfo {author} {\bibfnamefont {D.~P.}\ \bibnamefont
  {Kozlenko}}, \bibinfo {author} {\bibfnamefont {A.~A.}\ \bibnamefont {Belik}},
  \bibinfo {author} {\bibfnamefont {A.~V.}\ \bibnamefont {Belushkin}}, \bibinfo
  {author} {\bibfnamefont {E.~V.}\ \bibnamefont {Lukin}}, \bibinfo {author}
  {\bibfnamefont {W.~G.}\ \bibnamefont {Marshall}}, \bibinfo {author}
  {\bibfnamefont {B.~N.}\ \bibnamefont {Savenko}}, \ and\ \bibinfo {author}
  {\bibfnamefont {E.}~\bibnamefont {Takayama-Muromachi}},\ }\href {\doibase
  10.1103/PhysRevB.84.094108} {\bibfield  {journal} {\bibinfo  {journal} {Phys.
  Rev. B}\ }\textbf {\bibinfo {volume} {84}},\ \bibinfo {eid} {094108}
  (\bibinfo {year} {2011})}\BibitemShut {NoStop}%
\bibitem [{\citenamefont {Buhot}\ \emph {et~al.}(2015)\citenamefont {Buhot},
  \citenamefont {Toulouse}, \citenamefont {Gallais}, \citenamefont {Sacuto},
  \citenamefont {de~Sousa}, \citenamefont {Wang}, \citenamefont {Bellaiche},
  \citenamefont {Bibes}, \citenamefont {Barth{\'{e}}l{\'{e}}my}, \citenamefont
  {Forget}, \citenamefont {Colson}, \citenamefont {Cazayous},\ and\
  \citenamefont {Measson}}]{Buhot.2015}%
  \BibitemOpen
  \bibfield  {author} {\bibinfo {author} {\bibfnamefont {J.}~\bibnamefont
  {Buhot}}, \bibinfo {author} {\bibfnamefont {C.}~\bibnamefont {Toulouse}},
  \bibinfo {author} {\bibfnamefont {Y.}~\bibnamefont {Gallais}}, \bibinfo
  {author} {\bibfnamefont {A.}~\bibnamefont {Sacuto}}, \bibinfo {author}
  {\bibfnamefont {R.}~\bibnamefont {de~Sousa}}, \bibinfo {author}
  {\bibfnamefont {D.}~\bibnamefont {Wang}}, \bibinfo {author} {\bibfnamefont
  {L.}~\bibnamefont {Bellaiche}}, \bibinfo {author} {\bibfnamefont
  {M.}~\bibnamefont {Bibes}}, \bibinfo {author} {\bibfnamefont
  {A.}~\bibnamefont {Barth{\'{e}}l{\'{e}}my}}, \bibinfo {author} {\bibfnamefont
  {A.}~\bibnamefont {Forget}}, \bibinfo {author} {\bibfnamefont
  {D.}~\bibnamefont {Colson}}, \bibinfo {author} {\bibfnamefont
  {M.}~\bibnamefont {Cazayous}}, \ and\ \bibinfo {author} {\bibfnamefont
  {M.-A.}\ \bibnamefont {Measson}},\ }\href {\doibase
  10.1103/physrevlett.115.267204} {\bibfield  {journal} {\bibinfo  {journal}
  {Phys. Rev. Lett.}\ }\textbf {\bibinfo {volume} {115}},\ \bibinfo {pages}
  {267204} (\bibinfo {year} {2015})}\BibitemShut {NoStop}%
\bibitem [{\citenamefont {Haumont}\ \emph {et~al.}(2009)\citenamefont
  {Haumont}, \citenamefont {Bouvier}, \citenamefont {Pashkin}, \citenamefont
  {Rabia}, \citenamefont {Frank}, \citenamefont {Dkhil}, \citenamefont
  {Crichton}, \citenamefont {Kuntscher},\ and\ \citenamefont
  {Kreisel}}]{Haumont.2009}%
  \BibitemOpen
  \bibfield  {author} {\bibinfo {author} {\bibfnamefont {R.}~\bibnamefont
  {Haumont}}, \bibinfo {author} {\bibfnamefont {P.}~\bibnamefont {Bouvier}},
  \bibinfo {author} {\bibfnamefont {A.}~\bibnamefont {Pashkin}}, \bibinfo
  {author} {\bibfnamefont {K.}~\bibnamefont {Rabia}}, \bibinfo {author}
  {\bibfnamefont {S.}~\bibnamefont {Frank}}, \bibinfo {author} {\bibfnamefont
  {B.}~\bibnamefont {Dkhil}}, \bibinfo {author} {\bibfnamefont
  {W.}~\bibnamefont {Crichton}}, \bibinfo {author} {\bibfnamefont
  {C.}~\bibnamefont {Kuntscher}}, \ and\ \bibinfo {author} {\bibfnamefont
  {J.}~\bibnamefont {Kreisel}},\ }\href {\doibase 10.1103/PhysRevB.79.184110}
  {\bibfield  {journal} {\bibinfo  {journal} {Phys. Rev. B}\ }\textbf {\bibinfo
  {volume} {79}},\ \bibinfo {pages} {184110} (\bibinfo {year}
  {2009})}\BibitemShut {NoStop}%
\bibitem [{\citenamefont {Chen}\ \emph {et~al.}(2012)\citenamefont {Chen},
  \citenamefont {Lee}, \citenamefont {Huang},\ and\ \citenamefont
  {Lin}}]{Chen.2012(b)}%
  \BibitemOpen
  \bibfield  {author} {\bibinfo {author} {\bibfnamefont {Z.-W.}\ \bibnamefont
  {Chen}}, \bibinfo {author} {\bibfnamefont {J.-S.}\ \bibnamefont {Lee}},
  \bibinfo {author} {\bibfnamefont {T.}~\bibnamefont {Huang}}, \ and\ \bibinfo
  {author} {\bibfnamefont {C.-M.}\ \bibnamefont {Lin}},\ }\href {\doibase
  10.1016/j.ssc.2012.05.010} {\bibfield  {journal} {\bibinfo  {journal} {Solid
  State Commun.}\ }\textbf {\bibinfo {volume} {152}},\ \bibinfo {pages} {1613}
  (\bibinfo {year} {2012})}\BibitemShut {NoStop}%
\bibitem [{\citenamefont {Guennou}\ \emph
  {et~al.}(2011{\natexlab{b}})\citenamefont {Guennou}, \citenamefont {Bouvier},
  \citenamefont {Haumont}, \citenamefont {Garbarino},\ and\ \citenamefont
  {Kreisel}}]{Guennou.2011(b)}%
  \BibitemOpen
  \bibfield  {author} {\bibinfo {author} {\bibfnamefont {M.}~\bibnamefont
  {Guennou}}, \bibinfo {author} {\bibfnamefont {P.}~\bibnamefont {Bouvier}},
  \bibinfo {author} {\bibfnamefont {R.}~\bibnamefont {Haumont}}, \bibinfo
  {author} {\bibfnamefont {G.}~\bibnamefont {Garbarino}}, \ and\ \bibinfo
  {author} {\bibfnamefont {J.}~\bibnamefont {Kreisel}},\ }\href {\doibase
  10.1080/01411594.2011.552014} {\bibfield  {journal} {\bibinfo  {journal}
  {Phase Transit.}\ }\textbf {\bibinfo {volume} {84}},\ \bibinfo {pages} {474}
  (\bibinfo {year} {2011}{\natexlab{b}})}\BibitemShut {NoStop}%
\bibitem [{\citenamefont {Zhang}\ \emph {et~al.}(2013)\citenamefont {Zhang},
  \citenamefont {Wu}, \citenamefont {Zhang}, \citenamefont {Dong},
  \citenamefont {Wu}, \citenamefont {Liu}, \citenamefont {Wu},\ and\
  \citenamefont {Chen}}]{Zhang.2013}%
  \BibitemOpen
  \bibfield  {author} {\bibinfo {author} {\bibfnamefont {X.-L.}\ \bibnamefont
  {Zhang}}, \bibinfo {author} {\bibfnamefont {Y.}~\bibnamefont {Wu}}, \bibinfo
  {author} {\bibfnamefont {Q.}~\bibnamefont {Zhang}}, \bibinfo {author}
  {\bibfnamefont {J.-C.}\ \bibnamefont {Dong}}, \bibinfo {author}
  {\bibfnamefont {X.}~\bibnamefont {Wu}}, \bibinfo {author} {\bibfnamefont
  {J.}~\bibnamefont {Liu}}, \bibinfo {author} {\bibfnamefont {Z.-Y.}\
  \bibnamefont {Wu}}, \ and\ \bibinfo {author} {\bibfnamefont {D.-L.}\
  \bibnamefont {Chen}},\ }\href {\doibase 10.1088/1674-1137/37/12/128001}
  {\bibfield  {journal} {\bibinfo  {journal} {Chin. Phys. C}\ }\textbf
  {\bibinfo {volume} {37}},\ \bibinfo {pages} {128001} (\bibinfo {year}
  {2013})}\BibitemShut {NoStop}%
\bibitem [{\citenamefont {Guo}\ \emph {et~al.}(2019)\citenamefont {Guo},
  \citenamefont {Xing}, \citenamefont {Wang}, \citenamefont {Jia},
  \citenamefont {Zheng}, \citenamefont {Gong}, \citenamefont {Yang},
  \citenamefont {Li}, \citenamefont {Hao}, \citenamefont {Dong}, \citenamefont
  {Li}, \citenamefont {Li},\ and\ \citenamefont {Chen}}]{Guo.2019}%
  \BibitemOpen
  \bibfield  {author} {\bibinfo {author} {\bibfnamefont {Z.}~\bibnamefont
  {Guo}}, \bibinfo {author} {\bibfnamefont {H.}~\bibnamefont {Xing}}, \bibinfo
  {author} {\bibfnamefont {Y.}~\bibnamefont {Wang}}, \bibinfo {author}
  {\bibfnamefont {Q.}~\bibnamefont {Jia}}, \bibinfo {author} {\bibfnamefont
  {Z.}~\bibnamefont {Zheng}}, \bibinfo {author} {\bibfnamefont
  {Y.}~\bibnamefont {Gong}}, \bibinfo {author} {\bibfnamefont {D.}~\bibnamefont
  {Yang}}, \bibinfo {author} {\bibfnamefont {H.}~\bibnamefont {Li}}, \bibinfo
  {author} {\bibfnamefont {X.}~\bibnamefont {Hao}}, \bibinfo {author}
  {\bibfnamefont {J.}~\bibnamefont {Dong}}, \bibinfo {author} {\bibfnamefont
  {Y.}~\bibnamefont {Li}}, \bibinfo {author} {\bibfnamefont {X.}~\bibnamefont
  {Li}}, \ and\ \bibinfo {author} {\bibfnamefont {D.}~\bibnamefont {Chen}},\
  }\href {\doibase 10.1088/1361-648x/ab1469} {\bibfield  {journal} {\bibinfo
  {journal} {J. Phys.: Condens. Matter}\ }\textbf {\bibinfo {volume} {31}},\
  \bibinfo {pages} {265404} (\bibinfo {year} {2019})}\BibitemShut {NoStop}%
\bibitem [{\citenamefont {Shi}\ \emph {et~al.}(2016)\citenamefont {Shi},
  \citenamefont {Tang}, \citenamefont {Li}, \citenamefont {Liao}, \citenamefont
  {Wang}, \citenamefont {Ye},\ and\ \citenamefont {Xiong}}]{Shi.2016}%
  \BibitemOpen
  \bibfield  {author} {\bibinfo {author} {\bibfnamefont {P.-P.}\ \bibnamefont
  {Shi}}, \bibinfo {author} {\bibfnamefont {Y.-Y.}\ \bibnamefont {Tang}},
  \bibinfo {author} {\bibfnamefont {P.-F.}\ \bibnamefont {Li}}, \bibinfo
  {author} {\bibfnamefont {W.-Q.}\ \bibnamefont {Liao}}, \bibinfo {author}
  {\bibfnamefont {Z.-X.}\ \bibnamefont {Wang}}, \bibinfo {author}
  {\bibfnamefont {Q.}~\bibnamefont {Ye}}, \ and\ \bibinfo {author}
  {\bibfnamefont {R.-G.}\ \bibnamefont {Xiong}},\ }\href {\doibase
  10.1039/c5cs00308c} {\bibfield  {journal} {\bibinfo  {journal} {Chem. Soc.
  Rev.}\ }\textbf {\bibinfo {volume} {45}},\ \bibinfo {pages} {3811} (\bibinfo
  {year} {2016})}\BibitemShut {NoStop}%
\bibitem [{\citenamefont {Yang}\ \emph {et~al.}(2010)\citenamefont {Yang},
  \citenamefont {Seidel}, \citenamefont {Byrnes}, \citenamefont {Shafer},
  \citenamefont {Yang}, \citenamefont {Rossell}, \citenamefont {Yu},
  \citenamefont {Chu}, \citenamefont {Scott}, \citenamefont {Ager},
  \citenamefont {Martin},\ and\ \citenamefont {Ramesh}}]{Yang.2010}%
  \BibitemOpen
  \bibfield  {author} {\bibinfo {author} {\bibfnamefont {S.~Y.}\ \bibnamefont
  {Yang}}, \bibinfo {author} {\bibfnamefont {J.}~\bibnamefont {Seidel}},
  \bibinfo {author} {\bibfnamefont {S.~J.}\ \bibnamefont {Byrnes}}, \bibinfo
  {author} {\bibfnamefont {P.}~\bibnamefont {Shafer}}, \bibinfo {author}
  {\bibfnamefont {C.-H.}\ \bibnamefont {Yang}}, \bibinfo {author}
  {\bibfnamefont {M.~D.}\ \bibnamefont {Rossell}}, \bibinfo {author}
  {\bibfnamefont {P.}~\bibnamefont {Yu}}, \bibinfo {author} {\bibfnamefont
  {Y.-H.}\ \bibnamefont {Chu}}, \bibinfo {author} {\bibfnamefont {J.~F.}\
  \bibnamefont {Scott}}, \bibinfo {author} {\bibfnamefont {J.~W.}\ \bibnamefont
  {Ager}}, \bibinfo {author} {\bibfnamefont {L.~W.}\ \bibnamefont {Martin}}, \
  and\ \bibinfo {author} {\bibfnamefont {R.}~\bibnamefont {Ramesh}},\ }\href
  {\doibase 10.1038/nnano.2009.451} {\bibfield  {journal} {\bibinfo  {journal}
  {Nat. Nano}\ }\textbf {\bibinfo {volume} {5}},\ \bibinfo {pages} {143}
  (\bibinfo {year} {2010})}\BibitemShut {NoStop}%
\bibitem [{\citenamefont {Li}\ \emph {et~al.}(2018)\citenamefont {Li},
  \citenamefont {Lipatov}, \citenamefont {Lu}, \citenamefont {Lee},
  \citenamefont {Lee}, \citenamefont {Torun}, \citenamefont {Wirtz},
  \citenamefont {Eom}, \citenamefont {{\'{I}}{\~{n}}iguez}, \citenamefont
  {Sinitskii},\ and\ \citenamefont {Gruverman}}]{Li.2018}%
  \BibitemOpen
  \bibfield  {author} {\bibinfo {author} {\bibfnamefont {T.}~\bibnamefont
  {Li}}, \bibinfo {author} {\bibfnamefont {A.}~\bibnamefont {Lipatov}},
  \bibinfo {author} {\bibfnamefont {H.}~\bibnamefont {Lu}}, \bibinfo {author}
  {\bibfnamefont {H.}~\bibnamefont {Lee}}, \bibinfo {author} {\bibfnamefont
  {J.-W.}\ \bibnamefont {Lee}}, \bibinfo {author} {\bibfnamefont
  {E.}~\bibnamefont {Torun}}, \bibinfo {author} {\bibfnamefont
  {L.}~\bibnamefont {Wirtz}}, \bibinfo {author} {\bibfnamefont {C.-B.}\
  \bibnamefont {Eom}}, \bibinfo {author} {\bibfnamefont {J.}~\bibnamefont
  {{\'{I}}{\~{n}}iguez}}, \bibinfo {author} {\bibfnamefont {A.}~\bibnamefont
  {Sinitskii}}, \ and\ \bibinfo {author} {\bibfnamefont {A.}~\bibnamefont
  {Gruverman}},\ }\href {\doibase 10.1038/s41467-018-05640-4} {\bibfield
  {journal} {\bibinfo  {journal} {Nat. Commun.}\ }\textbf {\bibinfo {volume}
  {9}},\ \bibinfo {pages} {3344} (\bibinfo {year} {2018})}\BibitemShut
  {NoStop}%
\bibitem [{\citenamefont {Moubah}\ \emph {et~al.}(2012)\citenamefont {Moubah},
  \citenamefont {Rousseau}, \citenamefont {Colson}, \citenamefont {Artemenko},
  \citenamefont {Maglione},\ and\ \citenamefont {Viret}}]{Moubah.2012}%
  \BibitemOpen
  \bibfield  {author} {\bibinfo {author} {\bibfnamefont {R.}~\bibnamefont
  {Moubah}}, \bibinfo {author} {\bibfnamefont {O.}~\bibnamefont {Rousseau}},
  \bibinfo {author} {\bibfnamefont {D.}~\bibnamefont {Colson}}, \bibinfo
  {author} {\bibfnamefont {A.}~\bibnamefont {Artemenko}}, \bibinfo {author}
  {\bibfnamefont {M.}~\bibnamefont {Maglione}}, \ and\ \bibinfo {author}
  {\bibfnamefont {M.}~\bibnamefont {Viret}},\ }\href {\doibase
  10.1002/adfm.201201150} {\bibfield  {journal} {\bibinfo  {journal} {Adv.
  Funct. Mater.}\ }\textbf {\bibinfo {volume} {22}},\ \bibinfo {pages} {4814}
  (\bibinfo {year} {2012})}\BibitemShut {NoStop}%
\bibitem [{\citenamefont {Kreisel}, \citenamefont {Alexe},\ and\ \citenamefont
  {Thomas}(2012)}]{Kreisel.2012}%
  \BibitemOpen
  \bibfield  {author} {\bibinfo {author} {\bibfnamefont {J.}~\bibnamefont
  {Kreisel}}, \bibinfo {author} {\bibfnamefont {M.}~\bibnamefont {Alexe}}, \
  and\ \bibinfo {author} {\bibfnamefont {P.~A.}\ \bibnamefont {Thomas}},\
  }\href {\doibase 10.1038/nmat3282} {\bibfield  {journal} {\bibinfo  {journal}
  {Nat. Mater.}\ }\textbf {\bibinfo {volume} {11}},\ \bibinfo {pages} {260}
  (\bibinfo {year} {2012})}\BibitemShut {NoStop}%
\bibitem [{\citenamefont {Borkar}\ \emph {et~al.}(2018)\citenamefont {Borkar},
  \citenamefont {Rao}, \citenamefont {Tomar}, \citenamefont {Gupta},
  \citenamefont {Scott},\ and\ \citenamefont {Kumar}}]{Borkar.2018}%
  \BibitemOpen
  \bibfield  {author} {\bibinfo {author} {\bibfnamefont {H.}~\bibnamefont
  {Borkar}}, \bibinfo {author} {\bibfnamefont {V.}~\bibnamefont {Rao}},
  \bibinfo {author} {\bibfnamefont {M.}~\bibnamefont {Tomar}}, \bibinfo
  {author} {\bibfnamefont {V.}~\bibnamefont {Gupta}}, \bibinfo {author}
  {\bibfnamefont {J.}~\bibnamefont {Scott}}, \ and\ \bibinfo {author}
  {\bibfnamefont {A.}~\bibnamefont {Kumar}},\ }\href {\doibase
  10.1016/j.mtcomm.2017.12.004} {\bibfield  {journal} {\bibinfo  {journal}
  {Mater. Today Commun.}\ }\textbf {\bibinfo {volume} {14}},\ \bibinfo {pages}
  {116} (\bibinfo {year} {2018})}\BibitemShut {NoStop}%
\bibitem [{\citenamefont {Figielski}(1961)}]{Figielski.1961}%
  \BibitemOpen
  \bibfield  {author} {\bibinfo {author} {\bibfnamefont {T.}~\bibnamefont
  {Figielski}},\ }\href {\doibase 10.1002/pssb.19610010403} {\bibfield
  {journal} {\bibinfo  {journal} {Phys. Status Solidi B}\ }\textbf {\bibinfo
  {volume} {1}},\ \bibinfo {pages} {306} (\bibinfo {year} {1961})}\BibitemShut
  {NoStop}%
\bibitem [{\citenamefont {Wei}\ \emph {et~al.}(2017{\natexlab{a}})\citenamefont
  {Wei}, \citenamefont {Wang}, \citenamefont {Liu}, \citenamefont {Tsai},
  \citenamefont {Ke}, \citenamefont {Wu}, \citenamefont {Yin}, \citenamefont
  {Zhan}, \citenamefont {Lin}, \citenamefont {Chu},\ and\ \citenamefont
  {He}}]{Wei.2017(a)}%
  \BibitemOpen
  \bibfield  {author} {\bibinfo {author} {\bibfnamefont {T.-C.}\ \bibnamefont
  {Wei}}, \bibinfo {author} {\bibfnamefont {H.-P.}\ \bibnamefont {Wang}},
  \bibinfo {author} {\bibfnamefont {H.-J.}\ \bibnamefont {Liu}}, \bibinfo
  {author} {\bibfnamefont {D.-S.}\ \bibnamefont {Tsai}}, \bibinfo {author}
  {\bibfnamefont {J.-J.}\ \bibnamefont {Ke}}, \bibinfo {author} {\bibfnamefont
  {C.-L.}\ \bibnamefont {Wu}}, \bibinfo {author} {\bibfnamefont {Y.-P.}\
  \bibnamefont {Yin}}, \bibinfo {author} {\bibfnamefont {Q.}~\bibnamefont
  {Zhan}}, \bibinfo {author} {\bibfnamefont {G.-R.}\ \bibnamefont {Lin}},
  \bibinfo {author} {\bibfnamefont {Y.-H.}\ \bibnamefont {Chu}}, \ and\
  \bibinfo {author} {\bibfnamefont {J.-H.}\ \bibnamefont {He}},\ }\href
  {\doibase 10.1038/ncomms15108} {\bibfield  {journal} {\bibinfo  {journal}
  {Nat. Commun.}\ }\textbf {\bibinfo {volume} {8}},\ \bibinfo {pages} {15018}
  (\bibinfo {year} {2017}{\natexlab{a}})}\BibitemShut {NoStop}%
\bibitem [{\citenamefont {Kundys}\ \emph {et~al.}(2010)\citenamefont {Kundys},
  \citenamefont {Viret}, \citenamefont {Colson},\ and\ \citenamefont
  {Kundys}}]{Kundys.2010}%
  \BibitemOpen
  \bibfield  {author} {\bibinfo {author} {\bibfnamefont {B.}~\bibnamefont
  {Kundys}}, \bibinfo {author} {\bibfnamefont {M.}~\bibnamefont {Viret}},
  \bibinfo {author} {\bibfnamefont {D.}~\bibnamefont {Colson}}, \ and\ \bibinfo
  {author} {\bibfnamefont {D.~O.}\ \bibnamefont {Kundys}},\ }\href {\doibase
  10.1038/nmat2807} {\bibfield  {journal} {\bibinfo  {journal} {Nat. Mater.}\
  }\textbf {\bibinfo {volume} {9}},\ \bibinfo {pages} {803} (\bibinfo {year}
  {2010})}\BibitemShut {NoStop}%
\bibitem [{\citenamefont {Wei}\ \emph {et~al.}(2017{\natexlab{b}})\citenamefont
  {Wei}, \citenamefont {Wang}, \citenamefont {Li}, \citenamefont {Lin},
  \citenamefont {Hsieh}, \citenamefont {Chu},\ and\ \citenamefont
  {He}}]{Wei.2017(b)}%
  \BibitemOpen
  \bibfield  {author} {\bibinfo {author} {\bibfnamefont {T.-C.}\ \bibnamefont
  {Wei}}, \bibinfo {author} {\bibfnamefont {H.-P.}\ \bibnamefont {Wang}},
  \bibinfo {author} {\bibfnamefont {T.-Y.}\ \bibnamefont {Li}}, \bibinfo
  {author} {\bibfnamefont {C.-H.}\ \bibnamefont {Lin}}, \bibinfo {author}
  {\bibfnamefont {Y.-H.}\ \bibnamefont {Hsieh}}, \bibinfo {author}
  {\bibfnamefont {Y.-H.}\ \bibnamefont {Chu}}, \ and\ \bibinfo {author}
  {\bibfnamefont {J.-H.}\ \bibnamefont {He}},\ }\href {\doibase
  10.1002/adma.201701789} {\bibfield  {journal} {\bibinfo  {journal} {Adv.
  Mater.}\ }\textbf {\bibinfo {volume} {29}},\ \bibinfo {pages} {1701789}
  (\bibinfo {year} {2017}{\natexlab{b}})}\BibitemShut {NoStop}%
\bibitem [{\citenamefont {Uchino}\ and\ \citenamefont
  {Aizawa}(1985)}]{Uchino.1985}%
  \BibitemOpen
  \bibfield  {author} {\bibinfo {author} {\bibfnamefont {K.}~\bibnamefont
  {Uchino}}\ and\ \bibinfo {author} {\bibfnamefont {M.}~\bibnamefont
  {Aizawa}},\ }\href {\doibase 10.7567/jjaps.24s3.139} {\bibfield  {journal}
  {\bibinfo  {journal} {Jpn. J. Appl. Phys.}\ }\textbf {\bibinfo {volume}
  {24}},\ \bibinfo {pages} {139} (\bibinfo {year} {1985})}\BibitemShut
  {NoStop}%
\bibitem [{\citenamefont {Dingquan}\ \emph {et~al.}(1991)\citenamefont
  {Dingquan}, \citenamefont {Jianguo}, \citenamefont {Shipin}, \citenamefont
  {Xiu}, \citenamefont {Wen}, \citenamefont {Guoqin},\ and\ \citenamefont
  {Guanfeng}}]{Dingquan.1991}%
  \BibitemOpen
  \bibfield  {author} {\bibinfo {author} {\bibfnamefont {X.}~\bibnamefont
  {Dingquan}}, \bibinfo {author} {\bibfnamefont {Z.}~\bibnamefont {Jianguo}},
  \bibinfo {author} {\bibfnamefont {Z.}~\bibnamefont {Shipin}}, \bibinfo
  {author} {\bibfnamefont {W.}~\bibnamefont {Xiu}}, \bibinfo {author}
  {\bibfnamefont {Z.}~\bibnamefont {Wen}}, \bibinfo {author} {\bibfnamefont
  {L.}~\bibnamefont {Guoqin}}, \ and\ \bibinfo {author} {\bibfnamefont
  {X.}~\bibnamefont {Guanfeng}},\ }\href {\doibase
  10.1016/0038-1098(91)90460-d} {\bibfield  {journal} {\bibinfo  {journal}
  {Solid State Commun.}\ }\textbf {\bibinfo {volume} {79}},\ \bibinfo {pages}
  {1005} (\bibinfo {year} {1991})}\BibitemShut {NoStop}%
\bibitem [{\citenamefont {Schick}\ \emph {et~al.}(2014)\citenamefont {Schick},
  \citenamefont {Herzog}, \citenamefont {Wen}, \citenamefont {Chen},
  \citenamefont {Adamo}, \citenamefont {Gaal}, \citenamefont {Schlom},
  \citenamefont {Evans}, \citenamefont {Li},\ and\ \citenamefont
  {Bargheer}}]{Schick.2014}%
  \BibitemOpen
  \bibfield  {author} {\bibinfo {author} {\bibfnamefont {D.}~\bibnamefont
  {Schick}}, \bibinfo {author} {\bibfnamefont {M.}~\bibnamefont {Herzog}},
  \bibinfo {author} {\bibfnamefont {H.}~\bibnamefont {Wen}}, \bibinfo {author}
  {\bibfnamefont {P.}~\bibnamefont {Chen}}, \bibinfo {author} {\bibfnamefont
  {C.}~\bibnamefont {Adamo}}, \bibinfo {author} {\bibfnamefont
  {P.}~\bibnamefont {Gaal}}, \bibinfo {author} {\bibfnamefont {D.~G.}\
  \bibnamefont {Schlom}}, \bibinfo {author} {\bibfnamefont {P.~G.}\
  \bibnamefont {Evans}}, \bibinfo {author} {\bibfnamefont {Y.}~\bibnamefont
  {Li}}, \ and\ \bibinfo {author} {\bibfnamefont {M.}~\bibnamefont
  {Bargheer}},\ }\href {\doibase 10.1063/1.4901228} {\bibfield  {journal}
  {\bibinfo  {journal} {Phys. Rev. Lett.}\ }\textbf {\bibinfo {volume} {112}},\
  \bibinfo {pages} {097602} (\bibinfo {year} {2014})}\BibitemShut {NoStop}%
\bibitem [{\citenamefont {Burkert}, \citenamefont {Kreisel},\ and\
  \citenamefont {Kuntscher}(2016)}]{Burkert.2016}%
  \BibitemOpen
  \bibfield  {author} {\bibinfo {author} {\bibfnamefont {F.}~\bibnamefont
  {Burkert}}, \bibinfo {author} {\bibfnamefont {J.}~\bibnamefont {Kreisel}}, \
  and\ \bibinfo {author} {\bibfnamefont {C.}~\bibnamefont {Kuntscher}},\ }\href
  {\doibase 10.1063/1.4966548} {\bibfield  {journal} {\bibinfo  {journal}
  {Appl. Phys. Lett.}\ }\textbf {\bibinfo {volume} {109}},\ \bibinfo {pages}
  {182903} (\bibinfo {year} {2016})}\BibitemShut {NoStop}%
\bibitem [{\citenamefont {Meggle}\ \emph {et~al.}(2019)\citenamefont {Meggle},
  \citenamefont {Viret}, \citenamefont {Kreisel},\ and\ \citenamefont
  {Kuntscher}}]{Meggle.2019}%
  \BibitemOpen
  \bibfield  {author} {\bibinfo {author} {\bibfnamefont {F.}~\bibnamefont
  {Meggle}}, \bibinfo {author} {\bibfnamefont {M.}~\bibnamefont {Viret}},
  \bibinfo {author} {\bibfnamefont {J.}~\bibnamefont {Kreisel}}, \ and\
  \bibinfo {author} {\bibfnamefont {C.~A.}\ \bibnamefont {Kuntscher}},\ }\href
  {\doibase 10.1063/1.5081038} {\bibfield  {journal} {\bibinfo  {journal} {J.
  Appl. Phys.}\ }\textbf {\bibinfo {volume} {125}},\ \bibinfo {pages} {114104}
  (\bibinfo {year} {2019})}\BibitemShut {NoStop}%
\bibitem [{\citenamefont {Haumont}, \citenamefont {Saint-Martin},\ and\
  \citenamefont {Byl}(2008)}]{Haumont.2008}%
  \BibitemOpen
  \bibfield  {author} {\bibinfo {author} {\bibfnamefont {R.}~\bibnamefont
  {Haumont}}, \bibinfo {author} {\bibfnamefont {R.}~\bibnamefont
  {Saint-Martin}}, \ and\ \bibinfo {author} {\bibfnamefont {C.}~\bibnamefont
  {Byl}},\ }\href {\doibase 10.1080/01411590802328642} {\bibfield  {journal}
  {\bibinfo  {journal} {Phase Transit.}\ }\textbf {\bibinfo {volume} {81}},\
  \bibinfo {pages} {881} (\bibinfo {year} {2008})}\BibitemShut {NoStop}%
\bibitem [{\citenamefont {Klotz}\ \emph {et~al.}(2009)\citenamefont {Klotz},
  \citenamefont {Chervin}, \citenamefont {Munsch},\ and\ \citenamefont
  {Marchand}}]{Klotz.2009}%
  \BibitemOpen
  \bibfield  {author} {\bibinfo {author} {\bibfnamefont {S.}~\bibnamefont
  {Klotz}}, \bibinfo {author} {\bibfnamefont {J.-C.}\ \bibnamefont {Chervin}},
  \bibinfo {author} {\bibfnamefont {P.}~\bibnamefont {Munsch}}, \ and\ \bibinfo
  {author} {\bibfnamefont {G.~L.}\ \bibnamefont {Marchand}},\ }\href {\doibase
  10.1088/0022-3727/42/7/075413} {\bibfield  {journal} {\bibinfo  {journal} {J.
  Phys. D: Appl. Phys.}\ }\textbf {\bibinfo {volume} {42}},\ \bibinfo {pages}
  {075413} (\bibinfo {year} {2009})}\BibitemShut {NoStop}%
\bibitem [{\citenamefont {Syassen}(2008)}]{Syassen.2008}%
  \BibitemOpen
  \bibfield  {author} {\bibinfo {author} {\bibfnamefont {K.}~\bibnamefont
  {Syassen}},\ }\href {\doibase 10.1080/08957950802235640} {\bibfield
  {journal} {\bibinfo  {journal} {High Pressure Res.}\ }\textbf {\bibinfo
  {volume} {28}},\ \bibinfo {pages} {75} (\bibinfo {year} {2008})}\BibitemShut
  {NoStop}%
\bibitem [{\citenamefont {Himcinschi}\ \emph {et~al.}(2019)\citenamefont
  {Himcinschi}, \citenamefont {Rix}, \citenamefont {Röder}, \citenamefont
  {Rudolph}, \citenamefont {Yang}, \citenamefont {Rafaja}, \citenamefont
  {Kortus},\ and\ \citenamefont {Alexe}}]{Himcinschi.2019}%
  \BibitemOpen
  \bibfield  {author} {\bibinfo {author} {\bibfnamefont {C.}~\bibnamefont
  {Himcinschi}}, \bibinfo {author} {\bibfnamefont {J.}~\bibnamefont {Rix}},
  \bibinfo {author} {\bibfnamefont {C.}~\bibnamefont {Röder}}, \bibinfo
  {author} {\bibfnamefont {M.}~\bibnamefont {Rudolph}}, \bibinfo {author}
  {\bibfnamefont {M.-M.}\ \bibnamefont {Yang}}, \bibinfo {author}
  {\bibfnamefont {D.}~\bibnamefont {Rafaja}}, \bibinfo {author} {\bibfnamefont
  {J.}~\bibnamefont {Kortus}}, \ and\ \bibinfo {author} {\bibfnamefont
  {M.}~\bibnamefont {Alexe}},\ }\href {\doibase 10.1038/s41598-018-36462-5}
  {\bibfield  {journal} {\bibinfo  {journal} {Sci. Rep.}\ }\textbf {\bibinfo
  {volume} {9}},\ \bibinfo {pages} {379} (\bibinfo {year} {2019})}\BibitemShut
  {NoStop}%
\bibitem [{\citenamefont {Himcinschi}\ \emph {et~al.}(2015)\citenamefont
  {Himcinschi}, \citenamefont {Bhatnagar}, \citenamefont {Talkenberger},
  \citenamefont {Barchuk}, \citenamefont {Zahn}, \citenamefont {Rafaja},
  \citenamefont {Kortus},\ and\ \citenamefont {Alexe}}]{Himcinschi.2015}%
  \BibitemOpen
  \bibfield  {author} {\bibinfo {author} {\bibfnamefont {C.}~\bibnamefont
  {Himcinschi}}, \bibinfo {author} {\bibfnamefont {A.}~\bibnamefont
  {Bhatnagar}}, \bibinfo {author} {\bibfnamefont {A.}~\bibnamefont
  {Talkenberger}}, \bibinfo {author} {\bibfnamefont {M.}~\bibnamefont
  {Barchuk}}, \bibinfo {author} {\bibfnamefont {D.~R.~T.}\ \bibnamefont
  {Zahn}}, \bibinfo {author} {\bibfnamefont {D.}~\bibnamefont {Rafaja}},
  \bibinfo {author} {\bibfnamefont {J.}~\bibnamefont {Kortus}}, \ and\ \bibinfo
  {author} {\bibfnamefont {M.}~\bibnamefont {Alexe}},\ }\href {\doibase
  10.1063/1.4905443} {\bibfield  {journal} {\bibinfo  {journal} {Appl. Phys.
  Lett.}\ }\textbf {\bibinfo {volume} {106}},\ \bibinfo {pages} {012908}
  (\bibinfo {year} {2015})}\BibitemShut {NoStop}%
\bibitem [{\citenamefont {Liang}\ \emph {et~al.}(2019)\citenamefont {Liang},
  \citenamefont {Wang}, \citenamefont {Feng}, \citenamefont {Zhang},
  \citenamefont {Wang}, \citenamefont {Wang}, \citenamefont {Gu}, \citenamefont
  {Wu},\ and\ \citenamefont {Shen}}]{Liang.2019}%
  \BibitemOpen
  \bibfield  {author} {\bibinfo {author} {\bibfnamefont {Z.~W.}\ \bibnamefont
  {Liang}}, \bibinfo {author} {\bibfnamefont {Z.-H.}\ \bibnamefont {Wang}},
  \bibinfo {author} {\bibfnamefont {Y.}~\bibnamefont {Feng}}, \bibinfo {author}
  {\bibfnamefont {Q.~H.}\ \bibnamefont {Zhang}}, \bibinfo {author}
  {\bibfnamefont {L.~C.}\ \bibnamefont {Wang}}, \bibinfo {author}
  {\bibfnamefont {C.}~\bibnamefont {Wang}}, \bibinfo {author} {\bibfnamefont
  {L.}~\bibnamefont {Gu}}, \bibinfo {author} {\bibfnamefont {P.}~\bibnamefont
  {Wu}}, \ and\ \bibinfo {author} {\bibfnamefont {B.-G.}\ \bibnamefont
  {Shen}},\ }\href {\doibase 10.1103/physrevb.99.064304} {\bibfield  {journal}
  {\bibinfo  {journal} {Phys. Rev. B}\ }\textbf {\bibinfo {volume} {99}},\
  \bibinfo {pages} {064304} (\bibinfo {year} {2019})}\BibitemShut {NoStop}%
\bibitem [{\citenamefont {Kreisel}\ \emph {et~al.}(2011)\citenamefont
  {Kreisel}, \citenamefont {Jadhav}, \citenamefont {Chaix-Pluchery},
  \citenamefont {Varela}, \citenamefont {Dix}, \citenamefont {S{\'{a}}nchez},\
  and\ \citenamefont {Fontcuberta}}]{Kreisel.2011}%
  \BibitemOpen
  \bibfield  {author} {\bibinfo {author} {\bibfnamefont {J.}~\bibnamefont
  {Kreisel}}, \bibinfo {author} {\bibfnamefont {P.}~\bibnamefont {Jadhav}},
  \bibinfo {author} {\bibfnamefont {O.}~\bibnamefont {Chaix-Pluchery}},
  \bibinfo {author} {\bibfnamefont {M.}~\bibnamefont {Varela}}, \bibinfo
  {author} {\bibfnamefont {N.}~\bibnamefont {Dix}}, \bibinfo {author}
  {\bibfnamefont {F.}~\bibnamefont {S{\'{a}}nchez}}, \ and\ \bibinfo {author}
  {\bibfnamefont {J.}~\bibnamefont {Fontcuberta}},\ }\href {\doibase
  10.1088/0953-8984/23/34/342202} {\bibfield  {journal} {\bibinfo  {journal}
  {J. Phys.: Condens. Matter}\ }\textbf {\bibinfo {volume} {23}},\ \bibinfo
  {pages} {342202} (\bibinfo {year} {2011})}\BibitemShut {NoStop}%
\bibitem [{\citenamefont {Ramachandran}\ \emph {et~al.}(2010)\citenamefont
  {Ramachandran}, \citenamefont {Dixit}, \citenamefont {Naik}, \citenamefont
  {Lawes},\ and\ \citenamefont {Rao}}]{Ramachandran.2010}%
  \BibitemOpen
  \bibfield  {author} {\bibinfo {author} {\bibfnamefont {B.}~\bibnamefont
  {Ramachandran}}, \bibinfo {author} {\bibfnamefont {A.}~\bibnamefont {Dixit}},
  \bibinfo {author} {\bibfnamefont {R.}~\bibnamefont {Naik}}, \bibinfo {author}
  {\bibfnamefont {G.}~\bibnamefont {Lawes}}, \ and\ \bibinfo {author}
  {\bibfnamefont {M.~S.~R.}\ \bibnamefont {Rao}},\ }\href {\doibase
  10.1103/PhysRevB.82.012102} {\bibfield  {journal} {\bibinfo  {journal} {Phys.
  Rev. B}\ }\textbf {\bibinfo {volume} {82}},\ \bibinfo {pages} {012102}
  (\bibinfo {year} {2010})}\BibitemShut {NoStop}%
\bibitem [{\citenamefont {Xu}\ \emph {et~al.}(2009)\citenamefont {Xu},
  \citenamefont {Brinzari}, \citenamefont {Lee}, \citenamefont {Chu},
  \citenamefont {Martin}, \citenamefont {Kumar}, \citenamefont {McGill},
  \citenamefont {Rai}, \citenamefont {Ramesh}, \citenamefont {Gopalan},
  \citenamefont {Cheong},\ and\ \citenamefont {Musfeldt}}]{Xu.2009}%
  \BibitemOpen
  \bibfield  {author} {\bibinfo {author} {\bibfnamefont {X.~S.}\ \bibnamefont
  {Xu}}, \bibinfo {author} {\bibfnamefont {T.~V.}\ \bibnamefont {Brinzari}},
  \bibinfo {author} {\bibfnamefont {S.}~\bibnamefont {Lee}}, \bibinfo {author}
  {\bibfnamefont {Y.~H.}\ \bibnamefont {Chu}}, \bibinfo {author} {\bibfnamefont
  {L.~W.}\ \bibnamefont {Martin}}, \bibinfo {author} {\bibfnamefont
  {A.}~\bibnamefont {Kumar}}, \bibinfo {author} {\bibfnamefont
  {S.}~\bibnamefont {McGill}}, \bibinfo {author} {\bibfnamefont {R.~C.}\
  \bibnamefont {Rai}}, \bibinfo {author} {\bibfnamefont {R.}~\bibnamefont
  {Ramesh}}, \bibinfo {author} {\bibfnamefont {V.}~\bibnamefont {Gopalan}},
  \bibinfo {author} {\bibfnamefont {S.~W.}\ \bibnamefont {Cheong}}, \ and\
  \bibinfo {author} {\bibfnamefont {J.~L.}\ \bibnamefont {Musfeldt}},\ }\href
  {\doibase 10.1103/PhysRevB.79.134425} {\bibfield  {journal} {\bibinfo
  {journal} {Phys. Rev. B}\ }\textbf {\bibinfo {volume} {79}},\ \bibinfo {eid}
  {134425} (\bibinfo {year} {2009})}\BibitemShut {NoStop}%
\bibitem [{\citenamefont {Kumar}\ \emph {et~al.}(2008)\citenamefont {Kumar},
  \citenamefont {Rai}, \citenamefont {Podraza}, \citenamefont {Denev},
  \citenamefont {Ramirez}, \citenamefont {Chu}, \citenamefont {Martin},
  \citenamefont {Ihlefeld}, \citenamefont {Heeg}, \citenamefont {Schubert},
  \citenamefont {Schlom}, \citenamefont {Orenstein}, \citenamefont {Ramesh},
  \citenamefont {Collins}, \citenamefont {Musfeldt},\ and\ \citenamefont
  {Gopalan}}]{Kumar.2008}%
  \BibitemOpen
  \bibfield  {author} {\bibinfo {author} {\bibfnamefont {A.}~\bibnamefont
  {Kumar}}, \bibinfo {author} {\bibfnamefont {R.~C.}\ \bibnamefont {Rai}},
  \bibinfo {author} {\bibfnamefont {N.~J.}\ \bibnamefont {Podraza}}, \bibinfo
  {author} {\bibfnamefont {S.}~\bibnamefont {Denev}}, \bibinfo {author}
  {\bibfnamefont {M.}~\bibnamefont {Ramirez}}, \bibinfo {author} {\bibfnamefont
  {Y.-H.}\ \bibnamefont {Chu}}, \bibinfo {author} {\bibfnamefont {L.~W.}\
  \bibnamefont {Martin}}, \bibinfo {author} {\bibfnamefont {J.}~\bibnamefont
  {Ihlefeld}}, \bibinfo {author} {\bibfnamefont {T.}~\bibnamefont {Heeg}},
  \bibinfo {author} {\bibfnamefont {J.}~\bibnamefont {Schubert}}, \bibinfo
  {author} {\bibfnamefont {D.~G.}\ \bibnamefont {Schlom}}, \bibinfo {author}
  {\bibfnamefont {J.}~\bibnamefont {Orenstein}}, \bibinfo {author}
  {\bibfnamefont {R.}~\bibnamefont {Ramesh}}, \bibinfo {author} {\bibfnamefont
  {R.~W.}\ \bibnamefont {Collins}}, \bibinfo {author} {\bibfnamefont {J.~L.}\
  \bibnamefont {Musfeldt}}, \ and\ \bibinfo {author} {\bibfnamefont
  {V.}~\bibnamefont {Gopalan}},\ }\href {\doibase 10.1063/1.2901168} {\bibfield
   {journal} {\bibinfo  {journal} {Appl. Phys. Lett.}\ }\textbf {\bibinfo
  {volume} {92}},\ \bibinfo {pages} {121915} (\bibinfo {year}
  {2008})}\BibitemShut {NoStop}%
\bibitem [{Note1()}]{Note1}%
  \BibitemOpen
  \bibinfo {note} {As mentioned above, the energy of the absorption onset could
  also be estimated from the energy position of the Lorentzian used in the
  fitting of the absorbance spectrum. Using this energy position for
  calculating the energy difference $\Delta $ gives a similar pressure
  dependence as shown in Fig.\ \ref {fig:differeneces}, but with a constant
  offset.}\BibitemShut {Stop}%
\bibitem [{\citenamefont {Lobo}\ \emph {et~al.}(2007)\citenamefont {Lobo},
  \citenamefont {Moreira}, \citenamefont {Lebeugle},\ and\ \citenamefont
  {Colson}}]{Lobo.2007}%
  \BibitemOpen
  \bibfield  {author} {\bibinfo {author} {\bibfnamefont {R.}~\bibnamefont
  {Lobo}}, \bibinfo {author} {\bibfnamefont {R.}~\bibnamefont {Moreira}},
  \bibinfo {author} {\bibfnamefont {D.}~\bibnamefont {Lebeugle}}, \ and\
  \bibinfo {author} {\bibfnamefont {D.}~\bibnamefont {Colson}},\ }\href
  {\doibase 10.1103/PhysRevB.76.172105} {\bibfield  {journal} {\bibinfo
  {journal} {Phys. Rev. B}\ }\textbf {\bibinfo {volume} {76}},\ \bibinfo
  {pages} {172105} (\bibinfo {year} {2007})}\BibitemShut {NoStop}%
\end{thebibliography}
\end{document}